\documentclass[hidelinks]{aa}
\usepackage[varg]{txfonts}
\usepackage{bm}
\usepackage{enumitem,afterpage}
\usepackage{graphicx,tikz}
\usetikzlibrary{fadings,calc,decorations.pathmorphing,decorations.text}
\tikzset{snakearrow/.style={->,>=latex,decorate,decoration={snake,amplitude=.4mm,segment length=1mm,post length=1.4mm}}}
\tikzset{waveline/.style={decorate,decoration={snake,amplitude=0.5mm,segment length=8mm,pre length=2mm}}}
\usepackage{booktabs,multirow}
\usepackage[detect-all,range-units=single,range-phrase=\text{--},retain-zero-exponent=true]{siunitx}
\DeclareSIUnit\Gauss{G}
\DeclareSIUnit\erg{erg}
\DeclareSIUnit\sr{sr}
\DeclareSIUnit\Msol{\ensuremath{\mathrm{M_\odot}}}
\usepackage{natbib,twoopt}
\usepackage[breaklinks=true]{hyperref}
\bibpunct{(}{)}{;}{a}{}{,}
\defcitealias{NRF77}{Numerical Recipes}
\newcommand\nullfootnote[1]{%
	\begingroup%
	\renewcommand\thefootnote{}\footnote{#1}%
	\addtocounter{footnote}{-1}%
	\endgroup}
\makeatletter
\newcounter{subsubsubsection}[subsubsection]
\renewcommand\thesubsubsubsection{\thesubsubsection .\@arabic\c@subsubsubsection}
\newcommand\subsubsubsection{\@startsection{subsubsubsection}{4}{\z@}%
	{-3.25ex\@plus -1ex \@minus -.2ex}%
	{1.5ex \@plus .2ex}%
	{\normalfont\normalsize}}
\newcommand*\l@subsubsubsection{\@dottedtocline{4}{10.0em}{4.1em}}
\newcommand*{\subsubsubsectionmark}[1]{}
\providecommand{\toclevel@subsubsubsection}{4}
\makeatother
\newcommand{\referee}[1]{#1}

\begin{document}
	\title{New insight on Young Stellar Objects accretion shocks}
	\subtitle{--~a claim for NLTE opacities~--}

	\author {L. de Sá\inst{1,2}
		\and J.-P. Chièze$^\dagger$\inst{1,2}
		\and C. Stehlé\inst{1}
		\and I. Hubeny\inst{3}
		\and T. Lanz\inst{4}
		\and V. Cayatte\inst{5}}

	\institute{
			LERMA, Sorbonne Universités, Observatoire de Paris, PSL Research University, CNRS, F-75252, Paris, France\\
			\null\hfill\email{lionel.desa@obspm.fr}
		\and
			CEA/IRFU/SAp, CEA Saclay, Orme des Merisiers, 91191 Gif-sur-Yvette, France
		\and
			Steward Observatory, University of Arizona, 933 North Cherry Avenue, Tucson, AZ 85721, USA
		\and
			Observatoire de la Côte d’Azur (OCA), 06304 Nice, France
		\and
			LUTh, Observatoire de Paris, PSL University, CNRS, Paris-Diderot University, Meudon, France}

	\date{Started September 17, 2015}

	\abstract{Accreted material onto CTTSs is expected to form a hot quasi-periodic plasma structure that radiates in X-rays. Simulations of this phenomenon only partly match with observations. They all rely on a static model for the chromosphere model and on the assumption that radiation and matter are decoupled.}
			 {We explore the effects on the structure and on the dynamics of the accretion flow of both a shock-heated chromosphere and of the coupling between radiation and hydrodynamics.}
			 {We simulate accretion columns falling onto a stellar chromosphere using the 1D ALE code AstroLabE. This code solves the hydrodynamics equations along with the two first momenta equations for radiation transfer, with the help of a dedicated opacity table for the coupling between matter and radiation. We derive the total electron and ions densities from collisional-radiative NLTE ionisation equilibrium.}
			 {The chromospheric acoustic heating has an impact on the duration of the cycle and on the structure of the heated slab. In addition, the coupling between radiation and hydrodynamics leads to a heating of the accretion flow and the chromosphere, inducing a possible unburial of the whole column. These two last conclusions are in agreement with the computed monochromatic intensity. Both effects (acoustic heating and radiation coupling) have an influence on the amplitude and temporal variations of the net X-ray luminosity\referee{, which varies between 30 and 94\% of the incoming mechanical energy flux, depending on the model considered.}}{}

	\keywords{Stars: pre-main sequence -- Accretion, accretion disk -- Methods: numerical -- Hydrodynamics -- Radiative transfer -- Opacity}

	\maketitle

	\section{Introduction}


			Classical T Tauri Stars (CTTSs) are solar-type pre-main sequence stars surrounded by a thick disk composed of gas and dust \citep[see e.g.][]{Feigelson1999}. Disk material follows a near-Keplerian infall down to the \emph{truncation radius}, at which thermal and magnetic pressures balance. Free-falling material flows then from the inner disk down to the stellar surface in magnetically confined \emph{accretion columns} \citep{Calvet1998}. Hot spots observations \citep{Gullbring2000} suggest filling factors of up to 1\% \citep{Bouvier1995}.\nullfootnote{\hspace{-1em}\hspace{.2ex}$^\dagger$\hspace{.6ex} deceased.}

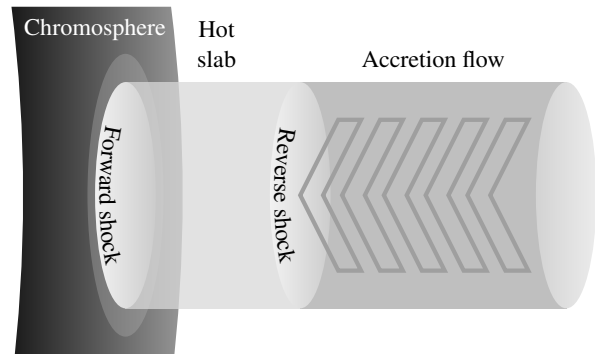
\begin{figure}[htb]
				\centering
				\begin{tikzpicture}
    \def\myshift#1{\raisebox{.4ex}}
    \clip(-.3,.4)rectangle(7.5,5);
    \shade[left color=black!90, middle color=black!85,right color=black!40](-.3,0) arc (-7:7:20.51)-- ++(2.1,0) arc (7:-7:20.51)--cycle;
    \fill[gray!60,opacity=.35](1.2,2.5)ellipse(5mm and 1.875cm);
    \fill[gray!30,opacity=.75](1.2,1)rectangle(3.5,4);
    \fill[gray!50](3.5,1)rectangle(7,4);
    \shade[left color=gray!10,right color=gray!40,postaction={decorate,decoration={text along path,text align=center,text={|\footnotesize\myshift|Forward shock}}}](1.2,2.5)ellipse(4mm and 1.5cm);
    \shade[left color=gray!10 ,right color=gray!40,postaction={decorate,decoration={text along path,text align=center,text={|\footnotesize\myshift|Reverse shock}}}](3.5,2.5)ellipse(4mm and 1.5cm);
    \shade[left color=gray!35,right color=gray!15](7,2.5)ellipse(4mm and 1.5cm);
    \foreach \i in {0,.55,...,2.3} {\draw[gray!75,ultra thick](4.3+\i,1.5)--(3.8+\i,2.5)--(4.3+\i,3.5)--(4+\i,3.5)--(3.5+\i,2.5)--(4+\i,1.5)--cycle;}
    \draw( .8 ,4.7)node{\footnotesize\color{white} Chromosphere};
    \draw(2.4 ,4.7)node{\footnotesize Hot};
    \draw(2.4 ,4.3)node{\footnotesize slab};
    \draw(5.25,4.3)node{\footnotesize Accretion flow};
\end{tikzpicture}
				\caption{Sketch of the basis of an accretion column and its three distinctive zones: the \emph{chromosphere} (\texttt{left, dark grey}), the \emph{accretion flow} (\texttt{right, mid-grey}) and the zone in between (\texttt{middle, light grey}) hereafter called \emph{hot slab} or \emph{post-shock medium}.\label{DESA_ref_Term}}
			\end{figure}

			Accreted gas is stopped where the flow ram pressure and the thermal pressure of the stellar chromosphere balance: a \emph{\referee{forward} shock} forms and the post-shock material accumulates at the basis of the column. The hot slab of post-shock material is separated from the accretion flow by a \emph{reverse shock}\footnote{The reverse shock is sometimes called \emph{accretion shock} in the literature.}. A typical simulated structure of an accretion shock can be found e.g. in \citet{Orlando2010} and is sketched in Figure \ref{DESA_ref_Term}.

			One of the most direct probes for the accretion process comes from the X-rays emitted by the dense ($n_\mathrm{e} > \SI{e11}{\per\cubic\cm}$) and hot ($T_\mathrm{\!e}\simeq\SIrange{2}{5}{\mega\K}$) post-shock plasma (see e.g. \citealp{Kastner2002} and \citealp{Stelzer2004} for TW Hya, \citealp{Schmitt2005} for BP Tau, \citealp{Gunther2006} for V4046 Sgr, \citealp{Argiroffi2007,Argiroffi2009} for MP Muscae, \citealp{Robrade2007} for RU Lup and \citealp{Huenemoerder2007} for Hen~3-600). Another signature is the UV-optical veiling, which is attributed to the post shock medium, the heated atmosphere and the pre-shock medium \citep{Calvet1998}. In addition, Doppler profiles of several emission lines trace the high velocity in the funneled flow \citep[up to \SI{500}{\km\per\s}, according to][]{Muzerolle1998}.\\

			1D hydrodynamical models \citep{Sacco2008,Sacco2010} predict Quasi-Periodic Oscillations (QPOs) of the post-shock slab with periods ranging from 0.01 to \SI{1000}{\s}, depending on the inflow density, metallicity, velocity and inclination with respect to the stellar surface. For a typical free-fall radial velocity of \SI{400}{\km\per\s}, \citet{Sacco2010} found for instance a period of \SI{160}{\s} at \SI{e11}{\per\cubic\cm}. These oscillations are triggered by the \emph{cooling instability} \citep[for further details, see e.g.][]{Chevalier1982,Walder1996,Mignone2005}.

			Although plasma characteristics derived from X-ray observations are consistent with the density and the temperature predicted by these numerical studies, there is no obvious observational evidence for such periodicity. \citet{Drake2009} studied thoroughly soft X-ray emission from TW Hydrae and found no periodicity in the range \SIrange{0.0001}{6.811}{\Hz}. \citet{Gunther2010} completed this study with optical and UV emission, and they came to the same conclusion in the range \SIrange{0.02}{50}{\Hz}. However, a recent photometric study of TW Hya based on MOST satellite observations reports possible oscillations with a period of \SIrange{650}{1200}{\s}, which could be assigned to post-shock plasma oscillations \citep{Siwak2018}.

			Observations thus raise the question of the existence of an oscillating hot slab in the accretion context. Several numerical studies explored multi-dimensional magnetic effects, like leaks at the basis of the column \citep{Orlando2010}, the tapering of the magnetic field \citep{Orlando2013}, or perturbations in the flow \citep{Matsakos2013}. Although QPOs are still obtained in these numerical studies, the accretion funnel basis is either fragmented in out-of-phase fibrils, or buried under a cooler and denser gas layer that strongly absorbs X-rays. The observation of global synchronous QPOs becomes therefore very challenging \citep{Curran2011,Bonito2014b,Colombo2016,Costa2017}. The effect of the slab burial into the chromosphere has also been explored in several 1D simulations \citep{Drake2005a,Sacco2010}. Depending on the depth of the burial, the radiation may only escape the post-shock structure from its upper part, leading to a significant reduction of the X-ray luminosity.\\


			In these numerical works, the accretion is supposed to take place on a quiet medium (an isothermal atmosphere in the best cases). Moreover, the post-shock medium is assumed to be optically thin, and the coupling between radiation and matter is reduced to a gas cooling function \citep[see e.g.][reported in Figure \ref{DESA_ref_Kirienko}]{Kirienko1993}. Although this assumption can be justified to model the infalling gas and the post-shock plasma, it is inconsistent with any stellar atmosphere model. The energy balance between radiation and gas in the lower stellar atmosphere is then replaced by a non-physical tuning (heating function, off threshold, \ldots). Such an assumption may affect the burial of the post-shock structure as well as the accretion structure itself.

			In this work, we focus and refine the physics encompassed in existing 1D models. We first explore the effect of chromospheric shocks perturbations on the accretion dynamics. We analyse then how radiation may affect the chromospheric, post-shock and accreted plasmas as well as the QPO duration and the hot slab burial; we also synthesise and discuss the accretion signature in the emerging radiative spectra. In Section \ref{DESA_ref_models}, we present the radiation hydrodynamics model and the numerical tools we use for the hydrodynamics and the spectra synthesis. We detail in Section \ref{DESA_ref_radsources} the two extreme radiative regimes encountered in this context, and a simple model for intermediate radiative regimes. Section \ref{DESA_ref_results} is dedicated to accretion simulations and to the corresponding discussions. The last section (Section \ref{DESA_ref_caveats}) presents caveats and possible improvements to this work.

	\section{Physical and numerical models\label{DESA_ref_models}}

		\subsection{Hydrodynamics model\label{DESA_ref_HD}}

			\subsubsection{Hydrodynamics equations\label{DESA_ref_HD_eqs}}

				We consider a star of radius $R\!_\star$ and mass $M\!_\star$. The accreted and stellar atmospheric plasmas at position $\vec{r}$ ($r = \left\lVert\vec{r}\right\lVert$), hereafter taken from the \emph{stellar surface}, are characterised by a (volumetric mass) density $\rho$, a velocity $\vec{v}$, a thermal pressure $p$ and a volumetric internal energy density $e$. The plasma evolution is modelled by solving the hydrodynamics equations, written in the conservative form:
				\begin{equation}\label{DESA_ref_eqHD}
					\left\{\begin{array}{@{\,}r@{\,}l@{}r@{\,}l@{\,}l}
						\partial_t\rho &+\vec{\nabla}\cdot&(\rho\,\vec{v}) &= 0\\[3pt]
						\partial_t(\rho\,\vec{v}) &+\vec{\nabla}&\hspace{-3pt}(\rho\,\vec{v}\otimes\vec{v}) &= \vec{\mathfrak{s}}_\mathrm{m} &= -\vec{\nabla}\left(p+p_\mathrm{vis}\right) + \vec{g}(\vec{r}) - \vec{\mathfrak{s}}_{M_\mathrm{r}}\\[3pt]
						\partial_t e &+\vec{\nabla}\cdot& (e\,\vec{v}) &= \mathfrak{s}_\mathrm{e} & = - \vec{\nabla}\cdot\left(p\,\vec{v}\right) + q_\mathrm{vis} - \vec{\nabla}\cdot \vec{q}_\mathrm{C} - \mathfrak{s}_{E_\mathrm{r}} - q_\chi
					\end{array}\right.\hspace{-2em}
				\end{equation}
				with $\vec{g}(\vec{r}) = - G M\!_\star \rho/\left(R\!_\star+r\right)^2\;\vec{r}/r$.

				The gas source terms\footnote{A sink is algebraically identified as \emph{negative} source term.} ($\mathfrak{s}_\mathrm{e}$ and $\vec{\mathfrak{s}}_\mathrm{m}$) include the contributions of thermal conduction \citep[$\vec{q}_\mathrm{C}$,][]{Spitzer1953,Vidal1995}, gravity ($\vec{g}(\vec{r})$), artificial viscosity \citep[$p_\mathrm{vis}$ and $q_\mathrm{vis}$,][]{vonNeumann1950} and the coupling with radiation ($\vec{\mathfrak{s}}_{M_\mathrm{r}}$ and $\mathfrak{s}_{E_\mathrm{r}}$, see Section \ref{DESA_ref_radsources}). The closure relation for this system of equations --~the equation of state~-- is adapted from the \emph{ideal gas law}: $p = n_\mathrm{tot}\,k\,T \Leftrightarrow e = 3/2\,p$, where $n_\mathrm{tot}$ stands for the total volumetric number density of free particles (neutrals, electrons and ions), and $T$ represents their kinetic temperature\footnote{All particles are assumed here to have the same kinetic temperature, i.e. $T_\mathrm{\!neutrals}=T_\mathrm{\!ions}=T_\mathrm{\!electrons}=T$.}. The contribution of ionisation/recombination on the gas energy density is included in the thermochemistry term $q_\chi$, and is discussed in the subsequent section (\ref{DESA_ref_ionisation}).

			\subsubsection{Collisional-radiative ionisation\label{DESA_ref_ionisation}}

				The \referee{forward} shock forms where the ram pressure is balanced by the local thermal pressure, i.e. within the stellar chromosphere, that needs then to be modelled. In contrary to the solar case, there is a very limited information about T Tauri chromospheres. Thus, as our goal is to propose a qualitative description of the dynamics of this chromosphere, and in absence of any reliable information, our chromospheric model (see Appendix \ref{DESA_ref_app_chromo}) is inspired by the solar case: therefore, we have chosen to use solar parameters in our simulations, and the chemical composition (solar abundances\footnote{Accreted material is expected to be depleted in heavy elements \citep{Fitzpatrick1996}. However, this phenomenon is not included in this study.}) is then taken from \citet{Grevesse1998}. In the hydrodynamics, we only consider hydrogen (\ion{H}{I}, \ion{H}{II}) and helium (\ion{He}{I}, \ion{He}{II}, \ion{He}{III}); the chemical composition is completed by a "catch-all" metal "M"\footnote{with a number abundance of 0.12\%, and a mass (averaged over abundances) of \SI{17}{\amu}.}.\\

				Most simulations are performed using time-independent ionisation models, for instance the modified Saha equilibrium of \citet{Brown1973} \citep[see e.g.][]{Sacco2008} or a detailed collisional ionisation calculation \citep[e.g.][]{Gunther2007}. To estimate the total free electron density $n_\mathrm{e}$ in the two first setups, we use the modified Saha model (for which $q_\chi=0$).\\
				The last simulation presented in this paper (referred to as the \emph{Hybrid} setup) uses a time-dependent \emph{collisional-radiative} ionisation model with:
				\begin{itemize}[nosep]
					\item \emph{collisional ionisation} rates given by \citet{Voronov1997};
					\item \emph{radiative recombination} rates computed by \citet{Verner1996a};
					\item helium \emph{dielectronic recombination} rate proposed by \citet{Hui1997};
					\item \emph{photo-ionisation} rates ($P$) derived from \citet{Spitzer1998} and \citet{Yan1998} cross-sections, and the local radiation energy density.
				\end{itemize}
				The time dependent ion and neutral volumetric number densities $n$ are then computed by a conservative set of equations (see e.g. \eqref{DESA_ref_eqHD}). The electron volumetric number density is then derived from the neutrality conservation: $n_\mathrm{e} = n_\ion{H}{II} + n_\ion{He}{II} + 2\,n_\ion{He}{III}$. Finally, the thermochemistry term $q_\chi$ sums all these contributions, weighted by the corresponding gained/lost energy.\\

				These calculations are performed independently from the opacity computation (see Appendix \ref{DESA_ref_app_opacities}), that uses a more refined version of the chemical composition \citep{Grevesse1998}.

		\subsection{Radiation model\label{DESA_ref_RT}}

			\subsubsection{Radiation and hydrodynamics}

				The coupling between radiation and matter enters at different scales in astrophysical plasmas. At a microscopic scale, radiation affects the thermodynamical state of the matter through its contribution to the populations of the electronic energy levels of each plasma ion. The computation of these populations is based on a large set of kinetic equilibrium equations that take into account excitation and de-excitation processes due to collisions (interactions with massive particles, mostly electrons) as well as radiative processes (interactions with photons). This step allows to derive also the monochromatic absorption and emission coefficients, resp. $\kappa_\nu$ (also called monochromatic \emph{opacity}, in \si{\square\cm\per\gram}) and $\eta_\nu$ (in \si{\erg\per\cubic\cm\per\s}), which in turn are used to compute the local radiation intensity by solving the equations of radiative transfer. Two limiting (and simplifying) cases are expected: at large electron densities, one recovers the Local Thermodynamic Equilibrium (LTE), whereas at low density and for an optically thin medium, the coronal limit is reached \citep{Oxenius1988}.

				The main issue in performing such calculations is an intricate coupling between the kinetic equilibrium equations (easily solved given the radiation field), and the radiative transfer equation (simple to calculate knowing the atomic level populations, and hence the absorption and emission coefficients). Since a mean free path of photons is typically much larger than the mean free path of massive particles, an explicit treatment of the radiation transport necessarily involves a significant non-locality of the problem. This issue is satisfactorily solved in the case of stationary stellar atmospheres \citep[see, e.g.][]{Hubeny2014}, using efficient iterative methods. However, this remains difficult in the case of a non-stationary plasma, where the equations of hydrodynamics need to be coupled, at each time, with the equations for the radiative transfer.

				Therefore, the previous kinetic equations have to be solved simultaneously with the monochromatic radiative transfer equations. This allows computing the frequency-averaged local radiation energy, flux and pressure, and helps including these quantities in the hydrodynamics equations (Eq. \eqref{DESA_ref_eqHD}). In practice, this exact description would require extensive numerical resources: the difficulty is commonly reduced by averaging the radiation quantities by frequency bands. In the \emph{multi-groups} approximation, the absorption and emission coefficients are averaged over several frequency bands using adapted weighting functions: the larger the number of groups, the better the precision of the computation. The simplest and most commonly used approach is the \emph{monogroup} approximation, which means that the radiation quantities are averaged over the whole frequency domain covered.

				Besides these delicate issues, radiative transfer takes part in the computation of the spectrum emerging from this structure. This is usually done by the post-processing of the hydrodynamic results by more detailed spectral synthesis tools, as detailed in Section \ref{DESA_ref_SYNSPEC}.

			\subsubsection{Moment equations\label{DESA_ref_RT_eqs}}

				The radiation field is described here by the \emph{momenta equations} \citep[see e.g.][]{Mihalas1984} for the frequency-integrated radiation energy volumetric density ($E_\mathrm{r}$, in \si{\erg\per\cubic\cm}) and momentum ($\vec{M}_\mathrm{\!r}$, in \si{\erg\per\cm\tothe{4}\s}) or flux\footnote{In our 1D hydrodynamics simulations, we only consider the component of vector quantities collinear to the accretion column.} (\mbox{$\vec{F}_\mathrm{\!r} = c^2\vec{M}_\mathrm{\!r}$}, in \si{\erg\per\square\cm\per\s}), written in the comoving frame \citep{Lowrie2001}:
				\begin{equation}\label{DESA_ref_eqRT}\newcommand{\scdot}{\!\cdot\!}
					\left\{\begin{array}{@{\,}l@{\,}l@{\,}l@{\,}l@{\,}l@{\,}r@{\,}l@{\,}l}
						\partial_t E_\mathrm{r} &+\,\vec{v}\scdot\partial_t\vec{M}_\mathrm{\!r} &+\,c^2&\vec{\nabla}\scdot\vec{M}_\mathrm{\!r} &+\left(\tens{P}_\mathrm{\!r}\!:\vec{\nabla}\right)\scdot&\vec{v} &+ \vec{\nabla}\scdot\left(E_\mathrm{r}\,\vec{v}\right) &= \mathfrak{s}_{E_\mathrm{r}}\\[5pt]
						\partial_t \vec{M}_\mathrm{\!r} &+\,\vec{v}\scdot\partial_t\tens{P}_\mathrm{\!r}/c^2 &+&\vec{\nabla}\scdot\tens{P}_\mathrm{\!r} &+\left(\vec{M}_\mathrm{\!r}\cdot\vec{\nabla}\right) &\vec{v} &+ \vec{\nabla} \left(\vec{M}_\mathrm{\!r}\scdot\vec{v}\right) &= \vec{\mathfrak{s}}_{M_\mathrm{r}}
					\end{array}\right.
				\end{equation}
				The (monogroup) radiation quantities are integrated from 1 to \SI{e4}{\angstrom}. The M1 closure relation allows then to derive the radiation pressure $\tens{P}_\mathrm{\!r}$ from the radiation energy density: $\tens{P}_\mathrm{\!r} = \tens{D}\,E_\mathrm{r}$. $\tens{D}$ and $\chi$ are respectively the Eddington tensor and factor ($\tens{D}\equiv\chi$ in 1D) and are defined as follows:
				\begin{equation}
					\tens{D} = \frac{1-\chi}{2}\tens{I_2} + \frac{3\chi-1}{2}\vec{i}\otimes\vec{i} \quad,\quad \chi = \frac{3+4f^2}{5+2\sqrt{4-3f^2}}
				\end{equation}
				with the reduced flux $\vec{f}=\vec{F}_\mathrm{\!r}/(c\,E_\mathrm{r})$ (and $f=\;\parallel\!\vec{f}\!\parallel$), the flux direction $\;\vec{i} = \vec{f}/f = \vec{F}_\mathrm{\!r}/F_\mathrm{\!r}$ and $\tens{I_2}$ the second-order identity tensor.\\
				\referee{As a drawback, the M1 radiation transfer may not properly model the radiation field in structures that involve more than one main radiation source \citep[see e.g.][]{Jiang2014b,Jiang2014a,Sadowski2014}. Moreover, }contrarily to the radiation energy, the contribution of the radiation flux to the hydrodynamics is not straightforward to interpret\footnote{For instance, in the case of an isotropic radiation, $\vec{F}_\mathrm{\!r}=\vec{0}$ whereas the radiation energy can be important.}; both are presented and discussed with our last setup (Section \ref{DESA_ref_ErFr}).

				Depending on the expression of the radiation source terms, these equations can continuously model optically thin to thick propagation media \citep[see e.g.][]{Mihalas1984}.

			\subsubsection{Radiation source terms - opacities \& line cooling\label{DESA_ref_radsources}}

				This work aims at describing in a consistent way the system composed of three zones, which are coupled together through radiation but in different thermodynamical states (Figure \ref{DESA_ref_Term}): the dense and optically thick \emph{near-LTE} chromosphere (Section \ref{DESA_ref_LTEtransfer}) on the one hand, the optically thin \emph{coronal} hot accretion slab and cold accretion flow (Section \ref{DESA_ref_Lambda}) on the other hand. We also expect, according e.g. to \citet{Calvet1998}, that the frequency distribution of the measured radiation varies strongly from the X-rays to the infrared. We have decided to work step by step, using a model which makes a continuous transition between the optically thick LTE approximation and the coronal limit, as described in Section \ref{DESA_ref_Zeta}.

				\subsubsubsection{Optically thick limit\label{DESA_ref_LTEtransfer}}

					The deep stellar atmosphere is optically thick and can be considered at LTE, i.e. each microphysics process is counter-balanced by its reverse process. In LTE and regimes close to LTE, the monochromatic absorption and emission coefficients are linked through the Planck distribution function: $\eta_\nu=\kappa_\nu\,\rho\,c\,B_\nu$. The radiation energy and momentum source terms are then defined by \citep[see e.g.][]{Mihalas1984}:
					\begin{equation}\label{DESA_ref_sourcesLTE}
						\mathfrak{s}_{E_\mathrm{r}}^* = \hphantom{-}\kappa_\mathrm{P}\,\rho\,c\left(a_\mathrm{R}\,T^4-E_\mathrm{r}\right)
						\qquad\text{and}\qquad
						\vec{\mathfrak{s}}_{M_\mathrm{r}}^* = -\kappa_\mathrm{R}\,\rho\,c\,\vec{M}_\mathrm{\!r}
					\end{equation}
					where $a_\mathrm{R}$ is the radiation constant. Two radiation-matter coupling factors appear here (in \si{\square\cm\per\gram}). The Planck mean opacity $\kappa_\mathrm{P}$ is based on the frequency-integrated absorption coefficient $\kappa_\nu$ weighted by the Planck distribution function $B_\nu$, while the Rosseland mean opacity $\kappa_\mathrm{R}$ is the harmonic mean of $\kappa_\nu$ weighted by the temperature derivative of the Planck function $\partial_T B_\nu$, as follows \citep{Mihalas1984}:
					\begin{equation}
						\kappa_\mathrm{P} = \frac{\int\kappa_\nu\,B_\nu\,d_{\!}\nu}{\int B_\nu\,d_{\!}\nu}
						\qquad\text{and}\qquad
						\kappa_\mathrm{R}^{-1} = \frac{\int\kappa_\nu^{-1}\,\partial_TB_\nu\,d_{\!}\nu}{\int \partial_TB_\nu\,d_{\!}\nu}
					\end{equation}
					In these frequency averages, the Planck mean is dominated by strong absorption features (typically lines), whereas the Rosseland mean is dominated by the regions in the spectrum of lowest monochromatic opacity. As a consequence, at large optical depths, $\kappa_\mathrm{P}$ correctly describes the energy exchange between particles and photons, while $\kappa_\mathrm{R}$ gives the correct total radiative flux \citep{Hubeny2014}.

					\begin{figure}[!ht]
						\centering
						\includegraphics[scale=.8]{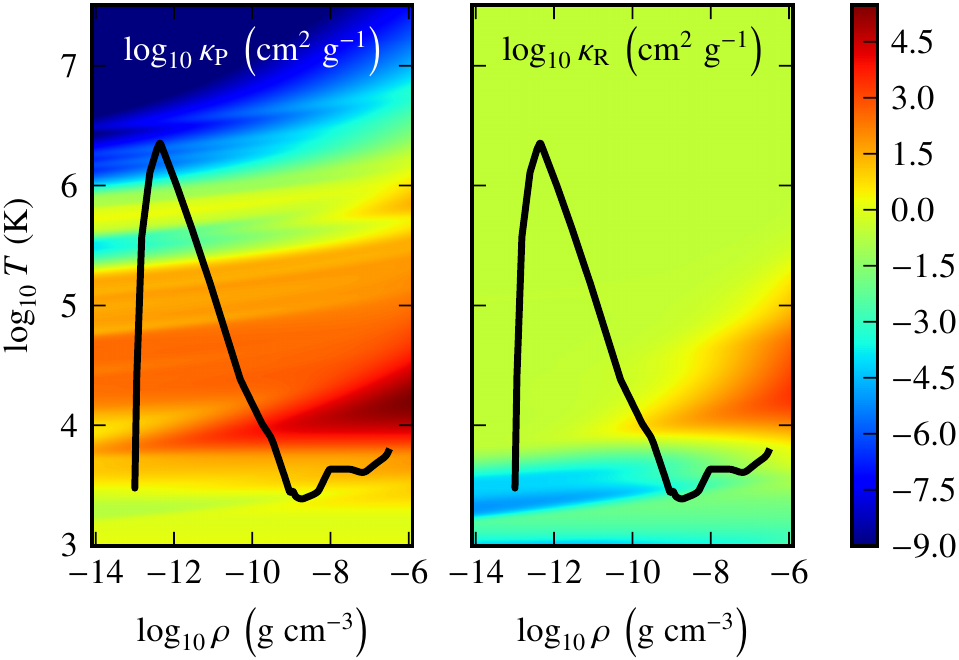}
						\caption{Planck ($\kappa_\mathrm{P}$, \texttt{left}) and Rosseland ($\kappa_\mathrm{R}$, \texttt{right}) opacities with respect to gas density and temperature, in log scale (cf. Appendix \ref{DESA_ref_app_opacities}). The \texttt{black curve} represents typical conditions met with chromosphere, accretion shock and flow.\label{DESA_ref_CSJF}}
					\end{figure}

					Several opacity tables are available for a variety of chemical compositions. However, they all fail to cover the full $(\rho,T)$ domain explored in our simulations (see solid black line in Figure \ref{DESA_ref_CSJF}). We constructed then with the SYNSPEC code (Section \ref{DESA_ref_SYNSPEC}) our own LTE opacity table (see Appendix \ref{DESA_ref_app_opacities} for further details), presented in Figure \ref{DESA_ref_CSJF}. These opacities include atomic (high $T$) and molecular (low $T$) contributions.

				\subsubsubsection{Optically thin limit\label{DESA_ref_Lambda}}

					Due to its very low density ($\rho\simeq\SI{e-13}{\g\per\cubic\cm}$), the accreted plasma can be described by the limit regime where the gas density tends towards zero: the \emph{coronal regime}. The coupling between radiation and matter boils down in this case to an optically thin radiative cooling function $\Lambda(T)$ (in \si{\erg\cubic\cm\per\s}). In Eq. \eqref{DESA_ref_eqRT}, the radiation source/sink terms become then:
					\begin{equation}\label{DESA_ref_sourcesCR}
						\mathfrak{s}^\dagger_{E_\mathrm{r}} = n_\mathrm{e}\,n_\mathrm{H}\,\Lambda(T)
						\qquad\text{and}\qquad
						\vec{\mathfrak{s}}^\dagger_{M_\mathrm{r}} = \vec{0}
					\end{equation}
					The first quantity represents the net radiation power emitted by unit volume in all directions ($\SI{4\pi}{\sr})$ by a hot optically thin plasma (in \si{\erg\per\cubic\cm\per\s\per\sr}). The term $\vec{\mathfrak{s}}^\dagger_{M_\mathrm{r}}$ is set to zero since there is no coupling between radiation and matter in this regime (see Appendix \ref{DESA_ref_app_radsourceshybrid} for more details).

					\begin{figure}[htb]
						\centering
						\includegraphics[width=.73\linewidth]{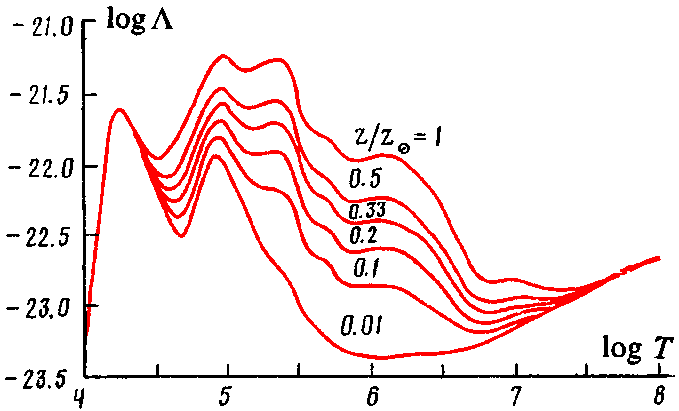}
						\caption{Optically thin radiative cooling (in \si{\erg\cubic\cm\per\s}) for different metallicities Z, versus gas temperature (\si{\K}), adapted from \citet{Kirienko1993}.\label{DESA_ref_Kirienko}}
					\end{figure}

					The present work is based on the cooling function provided by \citet{Kirienko1993}, reproduced in Figure \ref{DESA_ref_Kirienko}, with $Z/Z_\odot = 1$ (see Appendix \ref{DESA_ref_app_Motiv} for the explanation).

				\subsubsubsection{Intermediate regimes\label{DESA_ref_Zeta}}

					The previous source terms describe two well-defined plasma situations. On the one hand, the basis of the stellar chromosphere is optically thick and can be described by the previous LTE radiation source terms. On the other hand, the low density and hot slab is mostly optically thin and can be described in the coronal regime.\\
					It is physically expected and numerically compulsory to perform a smooth and continuous transition to encompass intermediate regimes. This could be done using adequate opacities and emissivities, as for instance obtained in a collisional-radiative model, unfortunately not available yet for the whole range of physical conditions of the present study.\\
					Thus we have preferred to follow the transition between LTE and coronal regimes with the probability for a photon (emitted from the column center) to escape sideways \citep[see e.g.][equation 3.66]{Lequeux2005}:
					\begin{equation}\label{DESA_ref_eqzeta}
						\zeta = \frac{1-\exp(-3\tau_\mathrm{e})}{3\tau_\mathrm{e}}\;,\quad \tau_\mathrm{e} = \kappa_\mathrm{P}\,\rho\,L_\mathrm{c}
					\end{equation}
					$\rho$ and $\kappa_\mathrm{P}$ values are taken at the photon emission position. The characteristic length $L_\mathrm{c}$ is here taken as the accretion column mean radius (i.e. \SI{1000}{\km}, see Section \ref{DESA_ref_strategy}). Radiation source terms become then (see Appendix \ref{DESA_ref_app_radsourceshybrid} for further details):
					\begin{equation}
						\mathfrak{s}_{E_\mathrm{r}} = (1-\zeta)\,\mathfrak{s}^*_{E_\mathrm{r}} + \zeta\mathfrak{s}^\dagger_{E_\mathrm{r}}
						\qquad\text{and}\qquad
						\vec{\mathfrak{s}}_{M_\mathrm{r}} = \vec{\mathfrak{s}}_{M_\mathrm{r}}^*
					\end{equation}
					the star ($^*$) and dagger ($\dagger$) denoting respectively the LTE (Eq. \eqref{DESA_ref_sourcesLTE}) and the coronal (Eq. \eqref{DESA_ref_sourcesCR}) expressions.

		\subsection{Numerical tools}

			\subsubsection{One-dimensional approach}

				Observations indicate that, in general, the ambient magnetic field is of the order of \SI{1}{\kilo\Gauss} \citep{Johns-Krull1999,Johns-Krull2007}. The resulting Larmor radius (\SI{1}{\mm}) is very small, i.e. the plasma follows the magnetic field lines. Moreover, the Alfvén velocity reaches 3\% of the speed of light and the magnetic waves behave thus like usual light waves. Therefore, focusing on the heart of an accretion column in strong magnetic field case, we can model the accreted material along one field line, that will be assumed to be radial relative to the stellar center. Since the accretion process is expected to involve strong shocks, we chose a numerical tool able to achieve very high spatial resolution.

			\subsubsection{AstroLabE -- an ALE code}

				The present work is based on numerical studies performed with the 1D Arbitrary-Lagrangian-Eulerian (ALE) code AstroLabE \citep[see e.g.][]{deSa2012,Chieze2012}. It is based on the Raphson-Newton solver \citepalias[Section 9]{NRF77} and a fully implicit scheme (the CFL condition can then be ignored) to compute primary variables at each time step.\\
				This code solves, along with the adequate physics and chemistry equations (see Sections \ref{DESA_ref_HD} and \ref{DESA_ref_RT}), the equations describing the behaviour of the grid points. The space discretisation can follow an Eulerian or a Lagrangian description. Moreover, the grid can freely adapt to hydrodynamics situations \citep[the \emph{arbitrary} description,][]{Dorfi1987}: this helps us reach high resolution around shocks with fixed cardinality ($\delta r/r_\mathrm{max}\simeq\num{e-7}$ with \mbox{150--300} grid points).\\
				Beside its application to stellar accretion \citep{deSaPhD2014}, AstroLabE has been used in several astrophysical situations such as the interstellar medium \citep{Lesaffre2002,Lesaffre2004}, experimental radiative shocks \citep{Stehle2002,Bouquet2004} or type Ia supernovae \citep{Charignon2013} studies.

			\subsubsection{SYNSPEC -- a spectrum synthesiser\label{DESA_ref_SYNSPEC}}

				For the computation of the opacities and of the emerging spectra, we used the public 1D spectrum synthesis code SYNSPEC \citep{Hubeny2017}. It is a multi-purpose code that can either construct a detailed synthetic spectrum for a given model atmosphere or disks, or generate LTE opacity tables. In this paper, we used SYNSPEC both for generating opacity tables (see Section \ref{DESA_ref_LTEtransfer} and Appendix \ref{DESA_ref_app_opacities}), and for the snapshots spectra presented in Section \ref{DESA_ref_spectra}.

				The resulting synthetic spectrum reflects the quality of the input astrophysical model; using an LTE model results in an LTE spectrum, while using a NLTE model results in a NLTE spectrum. The snapshots of our hydrodynamic simulations provide temperature and density as a function of position; it is therefore straightforward to compute LTE spectra for such structures. It would be in principle possible to construct approximate NLTE spectra, keeping temperature and density fixed from the hydrodynamic simulations (the so-called "restricted NLTE problem"). This could be done for instance by the computer program TLUSTY \citep{Hubeny1995,Hubeny2017}, which would provide NLTE level populations that can be communicated to SYNSPEC to produce detailed spectra. However, as previously mentioned, such a study is computationally very demanding and is well beyond the scope of the present paper. Nevertheless, since NLTE effects may be important, this will be done in a future paper. It will allow to inspect the effect of the LTE approximation on our results.

				This synthetic spectrum, computed at different altitudes of the accretion column, will reveal the role played by the different parts of the spectrum, from X-ray to Visible (\SIrange{1}{e4}{\angstrom}). However, it is important to note that, as the accretion column is limited in diameter, some effects, like the absorption by the coldest parts are only pertinent for an observation along or near the direction of the accretion column. A 3D radiative transfer post-processing would then be more suitable to the geometry of the system (\citealp{Ibgui2013}).

	\section{Accretion basis simulations\label{DESA_ref_results}}

		\subsection{Strategy and common parameters\label{DESA_ref_strategy}}

			We have simulated for this study several physical situations in order to check the net effect on the QPOs of the chromospheric model on one side and of the matter-radiation coupling on the other side. We present first the reference case: a gas flow hits a fixed, rigid and non-porous interface (W--$\Lambda$ case, Section \ref{DESA_ref_WL}). We check then the effect of a dynamically heated chromosphere on the accretion process (Chr--$\Lambda$ case, Section \ref{DESA_ref_CL}) and we finally check the effect of the radiation feedback on matter (Hybrid case, Section \ref{DESA_ref_Rad}). The conditions and main results of each simulation are resumed in Table \ref{DESA_ref_sumuptable}.\\

			\begin{table*}[htbp]
				\centering
				\caption{Characteristics of 3 simulations used in this work and their main results. "\texttt{W--$\mathtt{\Lambda}$}" corresponds to our \emph{reference} case.\label{DESA_ref_sumuptable}}
				\begin{tabular}{@{}rcccccccc@{}}
					\toprule
					\multirow{2}{*}{Name} & \multirow{2}{*}{\textbf{Atmos.}} & \textbf{Chromos.} & \textbf{Radiation} & \textbf{ionisation} &$\boldsymbol{H_\mathrm{max}}$ & $\boldsymbol{\tau_\mathrm{cycle}}$ & \multirow{2}{*}{\textbf{Section}} & \multirow{2}{*}{\textbf{Fig.}}\\
					& & \textbf{heating} & \textbf{source terms} & \textbf{model} & {\scriptsize($\times\num{e3}$ km)} & {\scriptsize(s)} &\\
					\cmidrule(r){1-1}\cmidrule(rl){2-5}\cmidrule(rl){6-7}\cmidrule(l){8-9}
					\textbf{W--$\boldsymbol{\Lambda}$} & "Window"* & -- & $\Lambda$* & Modified Saha* & 20 & 400 & \ref{DESA_ref_WL} & \ref{DESA_ref_ResWL}\\
					\cmidrule(r){1-1}\cmidrule(rl){2-5}\cmidrule(rl){6-7}\cmidrule(l){8-9}
					& \multirow{3}{*}{\parbox[c]{\widthof{Equilibrium}}{\centering Equilibrium atmosphere}} & $\mathcal{L}_\odot$* \& & \multirow{3}{*}{\parbox[c]{\widthof{\& $\Lambda$* (acc. flow)}}{\centering LTE (chromos.) \& $\Lambda$* (acc. flow)}}\\
					\textbf{Chr--$\boldsymbol{\Lambda}$} & & acoustic & & Modified Saha* & 17 & 350 & \ref{DESA_ref_CL} & \ref{DESA_ref_ResCL}\\
					& & heating\\
					\addlinespace
					\multirow{2}{*}{\textbf{Hybrid}} & Equilibrium & \multirow{2}{*}{$\mathcal{L}_\odot$*} & Intermediate & Time-dependent & \multirow{2}{*}{9} & \multirow{2}{*}{160} & \multirow{2}{*}{\ref{DESA_ref_Rad}} & \multirow{2}{*}{\ref{DESA_ref_ResH}}\\
					& atmosphere & & (transition: $\zeta$) & collisional radiative\\
					\bottomrule
				\end{tabular}
				\tablefoot{$H_\mathrm{max}$: maximum extension reached by the post-shock medium; $\mathtt{\tau}_\texttt{cycle}$: cycle duration; \texttt{"Window"}: fixed rigid non-porous transparent interface; $\mathtt{\Lambda}$: optically thin radiative cooling; $\mathcal{L}_\mathtt{\odot}$: one solar luminosity enters the simulation box from the inner boundary; \texttt{Modified Saha}: \citet{Brown1973}.}
			\end{table*}

			The simulations presented in this paper share few parameters:
			\begin{itemize}[nosep]
				\item the computational domain size is $r_\mathrm{out}=\SI{e5}{\km}$ (the outer boundary limit);
				\item the column\referee{/fibril} radius is set to\footnote{\referee{The ratio of the lateral to the longitudinal extension (in terms of typical radiative mean free path) of the column should be ideally large to justify 1D approximation for the computation of the effect of the radiative transfer throughout the system.}} $L_\mathrm{c} = \SI{1000}{\km}$\referee{, i.e. a filling factor of \num{2e-6}};
				\item for the gravity magnitude, we use $R\!_\star = R_\odot$ and $M\!_\star = M_\odot$;
				\item the accreted gas enters the computational domain through the outer boundary with $\rho_\mathrm{acc} = \SI{e-13}{\gram\per\cubic\cm}$, $T_\mathrm{\!acc} = \SI{3000}{\K}$~\footnote{In the Hybrid case, the temperature of the accretion flow is radiatively heated by the chromosphere up to $\SI{5730}{\K}$, before the accretion process starts.} and $v_\mathrm{acc} = \SI{400}{\km\per\s}$.
			\end{itemize}
			The velocity of the accreted gas is derived from the free-fall velocity at $r=r_\mathrm{out}$ above the stellar surface, considering a null radial velocity at the truncation radius $R_\mathrm{tr} = {\SI{2.2}{R_\odot}}$ (taken here from the center of the star).

			When the M1 radiation transfer is used (either near-LTE transfer or intermediate regime), one solar surface luminosity (\mbox{$\mathcal{L}_\odot = \SI{6.3e10}{\erg\per\square\cm\per\s}$}) enters from the inner boundary, and $c\times E_\mathrm{r}^\mathrm{out}/4$ leaves from the outer boundary\footnote{This expression is derived from the flux radiated outwards by an optically thin medium containing the radiation energy density $E_\mathrm{r}^\mathrm{out}$.}, with $E_\mathrm{r}^\mathrm{out}$ being the radiation energy density of the last computational cell.

		\subsection{Reference case (W--$\Lambda$)\label{DESA_ref_WL}}

			\subsubsection{Setup}

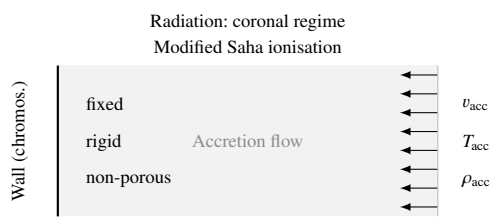
\begin{figure}[htb]
					\centering
					\begin{tikzpicture}
    \def\l{5}
    \fill[gray!10](0,0)rectangle(\l,2);
    \draw(-.5,1)node[text depth=0pt]{\rotatebox{90}{\scriptsize Wall (chromos.)}};
    \draw[gray!90](\l-2.5,1)node[text depth=0pt]{\scriptsize Accretion flow};
    \foreach\y in {0,.25,...,1.75} \draw[->,>=latex](\l,\y+.125)--(\l-.5,\y+.125);
    \draw[gray!50,thin](\l,0)--(\l,2);
    \draw[anchor=west](\l+.2,1.5)node[text depth=0pt]{\scriptsize $v_\mathrm{acc}$};
    \draw[anchor=west](\l+.2,1.0)node[text depth=0pt]{\scriptsize $T_\mathrm{\!acc}$};
    \draw[anchor=west](\l+.2,0.5)node[text depth=0pt]{\scriptsize $\rho_\mathrm{acc}$};
    \draw[thick](0,0)--(0,2);
    \draw[anchor=west](.25,1.5)node[text depth=0pt]{\scriptsize fixed};
    \draw[anchor=west](.25,1.0)node[text depth=0pt]{\scriptsize rigid};
    \draw[anchor=west](.25,0.5)node[text depth=0pt]{\scriptsize non-porous};
    \draw(\l/2,2.25)node[text depth=0pt]{\scriptsize Modified Saha ionisation};
    \draw(\l/2,2.6) node[text depth=0pt]{\scriptsize Radiation: coronal regime};
\end{tikzpicture}
					\caption{"W--$\Lambda$" simulation setup and boundary conditions.\label{DESA_ref_WLSetup}}
				\end{figure}

				In the reference case, we simulate the accretion stream using the same physics and assumptions than in previous models \citep[see e.g.][]{Sacco2008,Koldoba2008}. The matter-radiation coupling is then described by the coronal radiative cooling (Section \ref{DESA_ref_Lambda}) and the plasma ionisation is computed with the modified Saha equation (Section \ref{DESA_ref_ionisation}). In order to simplify the discussion, we focus on the post-shock structure and on the global dynamics. The stellar chromosphere is modelled in the simplest way, hereafter called the "window" model. It consists in a fixed rigid non-porous transparent interface. The main parameters are resumed in Figure \ref{DESA_ref_WLSetup}.

			\subsubsection{QPO cycle}

				\begin{figure*}[htbp]
					\centering
					\includegraphics[scale=.8]{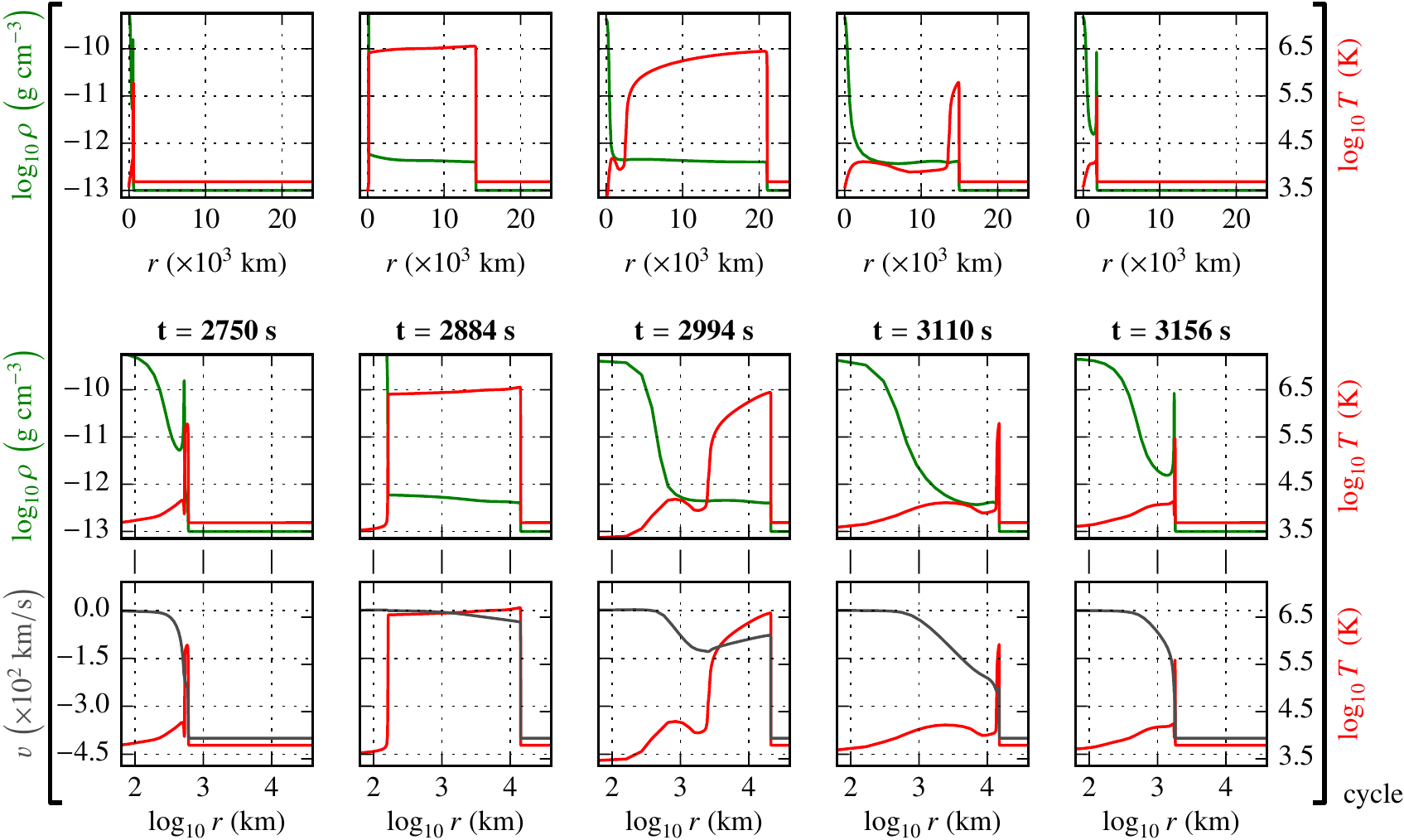}
					\caption{Lin-log (\texttt{top}) and log-log (\texttt{bottom}) snapshots of the density (\texttt{green}), temperature (\texttt{red}) and velocity (\texttt{grey}) profiles of a QPO cycle (\texttt{square brackets}) with the "W--$\Lambda$" setup; the accreted gas falls from the right to the left \citep[adapted from][]{deSa2014}.\newline
					\texttt{From left to right}: beginning of a new cycle ($t = \SI{2750}{\s}$), growth of a hot slab of shocked material ($t = \SI{2884}{\s}$), quasi-isochoric cooling at the slab basis (thermal instability, $t = \SI{2994}{\s}$), collapse of the post-shock structure (falling back of the reverse shock, $t = \SI{3110}{\s}$) and end of the collapse ($t = \SI{3156}{\s}$).\label{DESA_ref_ResWL}}
				\end{figure*}

				Besides the fact that matter accumulates on the left (inner) rigid boundary interface, the system is found to be perfectly periodic. Figure \ref{DESA_ref_ResWL} presents five snapshots of density, temperature and velocity profiles during a QPO cycle far from the initial stages. The accreted gas falls from right to left. A hot slab of shocked material builds first ($t = 2750$ and \SI{2884}{\s}) and cools down according to the coronal regime. Below a threshold temperature\footnote{i.e. the temperature at which the thermal instability is triggered ($\sim\SI{8e5}{\K}$) as expected from the optically thin radiative cooling variations with respect to temperature, see Section \ref{DESA_ref_Lambda} and references therein for further details.}, the fast, quasi-isochoric, cooling of the slab basis causes the collapse of the post-shock structure ($t = 2994$ and \SI{3110}{\s}). Just after the full collapse of the slab, since the accretion process is still working, a new slab forms and grows ($t = \SI{3156}{\s}$).

				This simulation is to be compared to the ones performed by \citet{Sacco2008}; Table \ref{DESA_ref_compSacco} resumes the main parameters and results for fast comparison. Despite few key differences (Sun vs. MP Muscæ parameters \& "window" vs. chromospheric heating function), the results are in good agreement with each other.

				\begin{table}
					\centering
					\caption{Comparison between our reference case ("W--$\Lambda$") and results obtained by \citet{Sacco2008}.\label{DESA_ref_compSacco}}
					\begin{tabular}{@{}r@{$\,$}lcc@{}}
						\toprule
						\multicolumn{2}{@{}r}{Parameters} & \textbf{\citeauthor{Sacco2008}} & \multirow{2}{*}{\textbf{"W--$\boldsymbol{\Lambda}$"}}\\
						\multicolumn{2}{@{}r}{\& quantities} & \textbf{(\citeyear{Sacco2008})}\\
						\cmidrule(r){1-2}\cmidrule(rl){3-3}\cmidrule(l){4-4}
						\multicolumn{2}{@{}r}{\bf Object} & MP Muscæ & Sun\\
						\addlinespace
						\multicolumn{2}{@{}r}{\bf Atmosphere} & Heating function & "Window"\\
						\addlinespace
						\multicolumn{2}{@{}r}{\bf Radiation} & $\Lambda$ & $\Lambda$\\
						\addlinespace
						\multicolumn{2}{@{}r}{\bf Ionisation} & Modified Saha & Modified Saha\\
						\addlinespace
						$\boldsymbol{\rho_\mathrm{acc}}$ & (g/cm$^3$) & \num{e-13} & \num{e-13}\\
						\addlinespace
						$\boldsymbol{v_\mathrm{acc}}$ & (km/s) & 450 & 400\\
						\addlinespace
						$\boldsymbol{T_\mathrm{acc}}$ & (\si{\K}) & \num{e3} & \num{3e3}\\
						\cmidrule(r){1-2}\cmidrule(rl){3-3}\cmidrule(l){4-4}
						$\boldsymbol{\tau_\mathrm{cycle}}$ & (\si{\s}) & 400 & 400\\
						\addlinespace
						$\boldsymbol{H_\mathrm{max}}$ & (Mm) & 18 & 20\\
						\addlinespace
						$\boldsymbol{n_\mathrm{e}}$ & (\si{\per\cubic\cm}) & \numrange{e11}{e12} & \numrange{e11}{e11.5}\\
						\addlinespace
						$\boldsymbol{T_\textbf{max}}$ & (\si{K}) & \num{e6.5} & \num{e6.5}\\
						\bottomrule
					\end{tabular}
				\end{table}

			\subsubsection{X-ray luminosity\label{DESA_ref_XRL}}

				An X-ray radiative power of \SI{1.3e30}{\erg\per\s} was measured in the range \SIrange{2}{27}{\angstrom} by \citet{Brickhouse2010} for TW Hydræ and an accretion flow velocity estimated at \SI{500}{\km\per\s}. We compute therefore the instantaneous X-ray surface luminosity $\mathcal{L}_{\!\Lambda}$ (in \si{\erg\per\square\cm\per\s}) and its time average $\bar{\mathcal{L}}_\mathrm{\!\Lambda}$\footnote{\referee{The "slab" is here defined as the plasma at temperature above \SI{E4.5}{\K}.}}:
				\begin{equation}\label{DESA_ref_eqbarF}
					\mathcal{L}_{\!\Lambda} = \int_\mathrm{slab}\hspace{-1em}n_\mathrm{e}\,n_\mathrm{H}\,\Lambda(T)\,d_{\!}r\quad\&\quad \bar{\mathcal{L}}_\mathrm{\!\Lambda} = \frac{1}{\tau_\mathrm{cycle}}\int_0^{\tau_\mathrm{cycle}} \hspace{-1.7em}\mathcal{L}_\mathrm{\!\Lambda}\,d_{\!}t
				\end{equation}
				to compare them with the values obtained with the different models presented in the subsequent sections and with the observational work of \citeauthor{Brickhouse2010}. These quantities are commonly compared to the incoming kinetic energy flux. However, since the flow \emph{accelerates} in its free-fall from the outer boundary down to the reverse shock, the plasma velocity and density may change between the outer boundary of the simulation box and the location of the reverse shock. To get round this issue, one must consider the \emph{mechanical} energy flux. This flux is calculated at any position $r$ by:
				\begin{equation}\label{DESA_ref_eqFM}
					F\!_\mathrm{M}= \frac{1}{2}\rho\,v(r)^3 + \int_{r_0}^r \hspace{-.3ex}\frac{G\,M\!_\star\,\rho\,v}{z^2}d_{\!}z
				\end{equation}
				where the origin of the gravitational energy potential is set at the mean \referee{forward} shock position ($r_0\simeq\SI{e3}{\km}$). The conservation of the mechanical energy induces that $F\!_\mathrm{M}$ does not depend on the position $r$. The value derived from our simulations is $F\!_\mathrm{M} = \SI{4.2e9}{\erg\per\square\cm\per\s}$.\\

				\begin{figure}[htb]
					\centering
					\includegraphics[scale=.8]{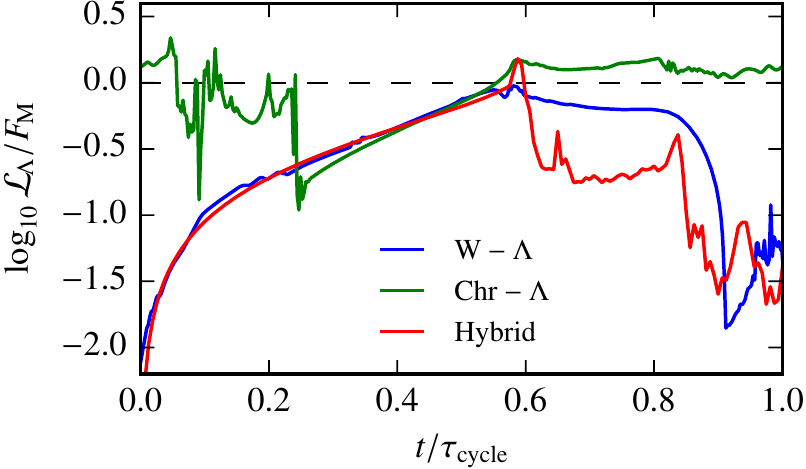}
					\caption{Time variation of the surface luminosity $\mathcal{L}_\Lambda$ \referee{emitted by} the hot slab for the \emph{reference (\texttt{blue})}, the \emph{dynamical chromosphere} (\texttt{green}) and the \emph{hybrid} (\texttt{red}) cases. These quantities are computed assuming an optically thin coronal plasma. To allow comparison, the time is reported in reduced units of $t/\tau_{cycle}$ and the luminosity is normalised to the incoming mechanical energy flux $F\!_\mathrm{M}$ defined in Eq. \eqref{DESA_ref_eqFM}. The values of the cycle duration for each setup is reported in Table \ref{DESA_ref_sumuptable}.\label{DESA_ref_Flux_Lambda}}
				\end{figure}

				Figure \ref{DESA_ref_Flux_Lambda} shows the time variation of $\mathcal{L}_{\!\Lambda}$. As expected, this quantity increases during the propagation of the reverse shock and decreases during the collapse. The time-averaged luminosity $\bar{\mathcal{L}}_{\!\Lambda} $ is equal to \SI{1.5e9}{\erg\per\square\cm\per\s}, i.e. $36\%$ of the incoming mechanical energy flux $F\!_\mathrm{M}$.

		\subsection{Effect of a dynamical chromosphere (Chr--$\Lambda$)\label{DESA_ref_CL}}

			\subsubsection{Setup\label{DESA_ref_CL-Setup}}

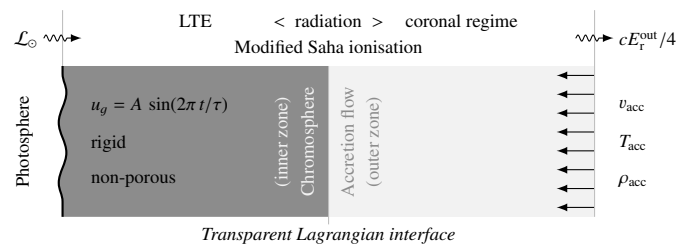
\begin{figure}[htb]
					\centering
					\begin{tikzpicture}
    \def\l{3.5}
    \fill[gray!10](0,0)rectangle(\l,2);
    \fill[gray!90]decorate[waveline]{(-\l,0)--(-\l,2)}--(0,2)--(0,0)--cycle;
    \draw(-\l-.5,1)node[text depth=0pt]{\rotatebox{90}{\scriptsize Photosphere}};
    \foreach\y in {0,.25,...,1.75} \draw[->,>=latex](\l,\y+.125)--(\l-.5,\y+.125);
    \draw[gray!50,thin](\l,0)--(\l,{2.6cm+4pt});
    \draw[snakearrow](\l-.25,{2.3cm+2pt})--(\l+.25,{2.3cm+2pt});\draw[anchor=west](\l+.2,{2.3cm+2pt})node[text depth=0pt]{\scriptsize $cE_\mathrm{r}^\mathrm{out}/4$};
    \draw[anchor=west](\l+.2,1.5)node[text depth=0pt]{\scriptsize $v_\mathrm{acc}$};
    \draw[anchor=west](\l+.2,1.0)node[text depth=0pt]{\scriptsize $T_\mathrm{\!acc}$};
    \draw[anchor=west](\l+.2,0.5)node[text depth=0pt]{\scriptsize $\rho_\mathrm{acc}$};
    \draw[gray!50,thin](0,-.1)--(0,2);
    \draw[gray!50,thin](-\l,2)--(-\l,{2.6cm+4pt});\draw[thick,waveline](-\l,0)--(-\l,2);
    \draw[snakearrow](-\l-.25,{2.3cm+2pt})--(-\l+.25,{2.3cm+2pt});\draw[anchor=east](-\l-.2,{2.3cm+2pt})node[text depth=0pt]{\scriptsize $\mathcal{L}_\odot$};
    \draw[anchor=west](-\l+.25,1.5)node[text depth=0pt]{\scriptsize $u_g = A\,\sin(2\pi\,t/\tau)$};
    \draw[anchor=west](-\l+.25,1.0)node[text depth=0pt]{\scriptsize rigid};
    \draw[anchor=west](-\l+.25,0.5)node[text depth=0pt]{\scriptsize non-porous};
    \draw(0,2.6) node[text depth=0pt]{\scriptsize $<\;$ radiation $\;>$};
    \draw(-\l/2,2.6)node[text depth=0pt]{\scriptsize LTE};
    \draw(\l/2,2.6)node[text depth=0pt]{\scriptsize coronal regime};
    \draw(0,2.25)node[text depth=0pt]{\scriptsize Modified Saha ionisation};
    \draw[gray!90]( .6 ,1)node[text depth=0pt]{\rotatebox{90}{\scriptsize (outer zone)}};
    \draw[gray!90]( .25,1)node[text depth=0pt]{\rotatebox{90}{\scriptsize Accretion flow}};
    \draw[gray!10](-.6 ,1)node[text depth=0pt]{\rotatebox{90}{\scriptsize (inner zone)}};
    \draw[gray!10](-.25,1)node[text depth=0pt]{\rotatebox{90}{\scriptsize Chromosphere}};
    \draw(0,-.25)node{\scriptsize\it Transparent Lagrangian interface};
\end{tikzpicture}
\vspace{-.5\baselineskip}
					\caption{"Chr--$\Lambda$" simulation setup and boundary conditions. \mbox{$A = \SI{0.6575}{\km\per\s}$} and $\tau = \SI{60}{\s}$.\label{DESA_ref_CLSetup}}
				\end{figure}

				In this second setup (see Figure \ref{DESA_ref_CLSetup}), we aim at studying the effect of a dynamically heated chromosphere on the phenomenon described in the previous Section. To achieve this, we "divide" the computational domain into two zones separated by a transparent\footnote{Although the column plasma is expected to be at coronal regime, LTE radiation transfer is needed to build the chromosphere layer. It is therefore essential to allow radiation to escape from the first zone through the second (optically thin) one.} Lagrangian interface.\\
				The outer zone is described as before, i.e. with modified Saha ionisation and optically thin radiative cooling (coronal regime). However, the inner zone is now described by our chromospheric model (see Appendix \ref{DESA_ref_app_chromo}). Ionisation is still described by the modified Saha equation, but we use the LTE radiation source terms as given in Eq. \eqref{DESA_ref_sourcesLTE}. To get a dynamically heated chromosphere, we first compute a \referee{radiative-}hydrostatic equilibrium, with the outer zone inactivated, and with one solar luminosity crossing the entire domain (no effect on the outer zone). Acoustic energy is then injected in the form of monochromatic sinusoidal motion of the first interface (a "window") with a \SI{60}{\s} period to mimic solar granulation. Several snapshots of temperature profiles are presented in Figure \ref{DESA_ref_app_chromo-snap}. Once the shock-heated chromosphere reaches its stationary regime, the accretion process is launched (in the outer zone).

			\subsubsection{Acoustic perturbations}

				\begin{figure*}[htb]
					\centering
					\includegraphics[scale=.8]{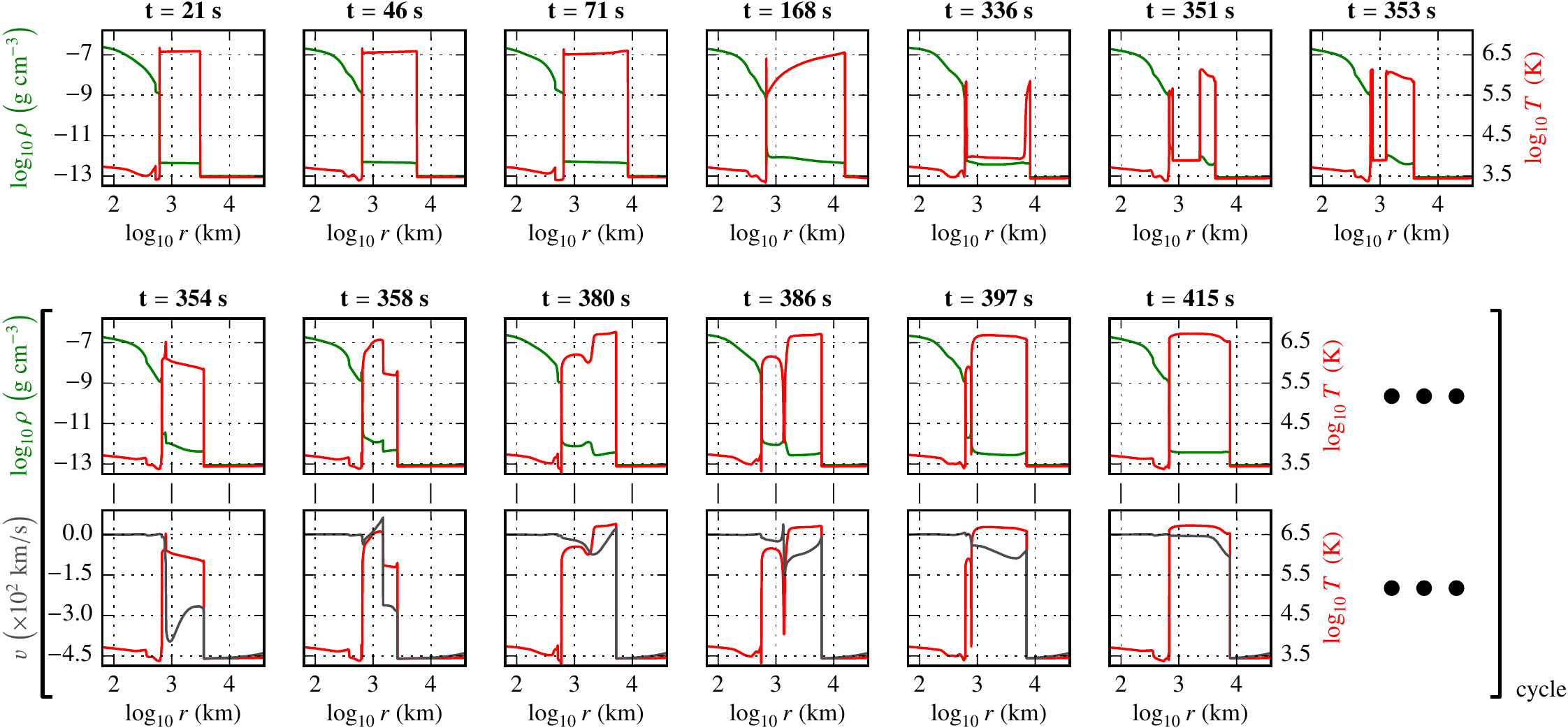}
					\caption{Snapshots of the density (\texttt{green}), temperature (\texttt{red}) and gas velocity (\texttt{grey}) profiles of the first QPO cycle with the "Chr--$\Lambda$" setup; the accreted gas falls from the right to the left. The first line (between $1$ and \SI{353}{\s}) corresponds to the first cycle. The second and third lines correspond to the beginning of the second cycle. Snapshots at $t = \SI{71}{\s}$ and \SI{415}{\s} are very close: from this time, the cycle behaves like the previous one. A typical sequence is: growth of a hot slab of shocked material ($t = \SI{21}{\s}$), quasi-isochoric cooling at the slab basis (thermal instability, $t = \SI{168}{\s}$), start of the collapse of the post-shock structure ($t = \SI{336}{\s}$), impact of the collapsing material on the chromosphere ($t=\SI{354}{\s}$), launch of a new shock \emph{before} the end of the collapse ($t = \SI{358}{\s}$), passing of the two shocks ($t = \SI{380}{\s}$), end of the collapse of the "old" structure ($t = \SI{386}{\s}$) and growth of the new slab ($t = \SI{415}{\s}$).\label{DESA_ref_ResCL}}
				\end{figure*}

				Figure \ref{DESA_ref_ResCL} shows seven snapshots of density and temperature profiles during the first QPO cycle (\SIrange{1}{354}{\s}). They are followed in the second line by 5 snapshots of the second QPO cycle (\SIrange{354}{415}{\s}). The second cycle differs from the first one only during the slab building (\SIrange{354}{397}{\s}). The sixth snapshot (\SI{415}{\s}) is very close to the snapshot of the first cycle at $t=\SI{71}{\s}$. The (unchanged) end of the second cycle is then not reported.

				During the installation phase (\SIrange{1}{336}{\s}) of the reverse shock, the post-shock structure follows more or less the same scenario than for the reference case (W--$\Lambda$). After several periods of the acoustic waves, small differences occur. The transmission of these waves/shocks to the accretion column depends on the leap of the acoustic impedance between the upper chromosphere and the hot slab, which results in reflection/transmission of these waves/shocks at this interface. The smallest leap is reached at the end of the collapse, near \SI{336}{\s}, leading to a transmission increase, which however remains still low. Their effect leads to small perturbations in the post-shock density (as it can already be seen at \SI{168}{\s}).

				\begin{table}[htb]
					\centering
					\caption{Position of the \emph{old} and \emph{new} reverse shocks between $t = \SI{358}{\s}$ and $t = \SI{397}{\s}$ (see Figure \ref{DESA_ref_ResCL}).\label{DESA_ref_shocktable}}
					\begin{tabular}{@{}rcccc@{}}
						\toprule
						Time (\si{s}) & \textbf{358} & \textbf{380} & \textbf{386} & \textbf{397}\\
						\cmidrule(r){1-1}\cmidrule(l){2-5}
						$\boldsymbol{r_\mathrm{old}}$ \textbf{(\si{km})} & \num{e3.40} & \num{e3.25} & \num{e3.10} & \num{e2.95}\\
						$\boldsymbol{r_\mathrm{new}}$ \textbf{(\si{km})} & \num{e3.15} & \num{e3.70} & \num{e3.80} & \num{e3.90}\\
						\bottomrule
					\end{tabular}
				\end{table}

				After this time, the transmitted waves start to feed with matter the hot collapsing layer behind the reverse shock. The thickness of this layer increases, as can be shown in Figure \ref{DESA_ref_ResCL} at \SI{351}{\s}, compared for instance with our reference case (\SI{3110}{\s}, Figure \ref{DESA_ref_ResWL}). 
				This structure collapses and hits at \SI{354}{\s} the dense chromosphere, leading to a secondary reverse shock which propagates backwards inside the slab. This behaviour is confirmed by the velocity variations shown in grey in Figure \ref{DESA_ref_ResCL}. The two reverse shocks pass then each other: the positions of the \emph{new} shock (or contact discontinuity) and the previous (\emph{old}) one are resumed in Table \ref{DESA_ref_shocktable}. The end of one cycle therefore overlaps the beginning of a new one.

			\subsubsection{Observational consequences\label{DESA_ref_CLObs}}

				This model implies two main observational consequences. First, compared to the reference case, the QPO cycle period is modified by the acoustic heating. The question of possible resonance is pointless regarding multi-mode acoustic heating by out-of-phase waves emitted in different locations. The period $\tau_\mathrm{cycle}$ is slightly reduced (from \SI{400}{\s} for the W--$\Lambda$ model to \SI{350}{\s} here, Table \ref{DESA_ref_sumuptable}) when using solar chromospheric parameters. Since CTTSs' atmospheres have a stronger activity than the Sun's one (that we use for the chromospheric model), the effect is expected to be enhanced in CTTSs.\\
				The second effect deals with the X-ray luminosity variation during a cycle, as reported in green in Figure \ref{DESA_ref_Flux_Lambda}. The growth phase is comparable with the W--$\Lambda$ setup, but the acoustic perturbations from the chromosphere induce strong differences in the collapse phase. Moreover, the overlapping of the beginning and end of the cycles affect the X-ray luminosity and the overall amplitude of the variations (contrast) is reduced compared to the reference case. QPO observations may thus require both higher time resolution and improved sensitivity. The time averaged surface luminosity (Eq. \eqref{DESA_ref_eqbarF}) is here equal to \SI{4.0e9}{\erg\per\square\cm\per\s}, i.e. $94\%$ of the mechanical energy flux $F_\mathrm{\!M}$.\\

				These results show that, compared to the reference case, the dynamical heating of the chromosphere impacts the duration of the QPO period and its observability. Of course, a more realistic description of the chromospheric heating would require at least a 2D MHD picture. For instance, we know that chromospheric perturbations may lead -- inside the column -- to the development of fibrils \citep[see e.g.][ChrFlx\# models]{Matsakos2013}, which is one of the scenarii explaining the absence of observation of QPO. In the acoustic description of the chromospheric heating, these fibrils, evolving out of phase, will also be strongly affected by the chromospheric perturbations.

		\subsection{Radiation effect on accretion (Hybrid)\label{DESA_ref_Rad}}

			\subsubsection{Setup}

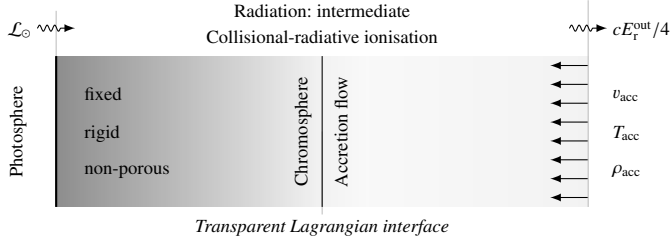
\begin{figure}[htb]
					\centering
					\begin{tikzpicture}
    \def\l{3.5}
    \shade[left color=gray!90,middle color=gray!80,right color=gray!10](-\l,0)rectangle(0,2);
    \shade[left color=gray!10,right color=gray!5](0,0)rectangle(\l/6,2);
    \shade[left color=gray!5,middle color=gray!10,right color=gray!10](\l/6,0)rectangle(\l,2);
    \draw(-\l-.5,1)node[text depth=0pt]{\rotatebox{90}{\scriptsize Photosphere}};
    \foreach\y in {0,.25,...,1.75} \draw[->,>=latex](\l,\y+.125)--(\l-.5,\y+.125);
    \draw[gray!50,thin](\l,0)--(\l,{2.6cm+4pt});
    \draw[snakearrow](\l-.25,{2.3cm+2pt})--(\l+.25,{2.3cm+2pt});\draw[anchor=west](\l+.2,{2.3cm+2pt})node[text depth=0pt]{\scriptsize $cE_\mathrm{r}^\mathrm{out}/4$};
    \draw[anchor=west](\l+.2,1.5)node[text depth=0pt]{\scriptsize $v_\mathrm{acc}$};
    \draw[anchor=west](\l+.2,1.0)node[text depth=0pt]{\scriptsize $T_\mathrm{\!acc}$};
    \draw[anchor=west](\l+.2,0.5)node[text depth=0pt]{\scriptsize $\rho_\mathrm{acc}$};
    \draw[gray!50,thin](0,-.1)--(0,1);
    \draw[thin](0,0)--(0,2);
    \draw[gray!50,thin](-\l,2)--(-\l,2.6);\draw[thick](-\l,0)--(-\l,2);
    \draw[snakearrow](-\l-.25,{2.3cm+2pt})--(-\l+.25,{2.3cm+2pt});\draw[anchor=east](-\l-.2,{2.3cm+2pt})node[text depth=0pt]{\scriptsize $\mathcal{L}_\odot$};
    \draw[anchor=west](-\l+.25,1.5)node[text depth=0pt]{\scriptsize fixed};
    \draw[anchor=west](-\l+.25,1.0)node[text depth=0pt]{\scriptsize rigid};
    \draw[anchor=west](-\l+.25,0.5)node[text depth=0pt]{\scriptsize non-porous};
    \draw(0,2.25)node[text depth=0pt]{\scriptsize Collisional-radiative ionisation};
    \draw(0,2.6) node[text depth=0pt]{\scriptsize Radiation: intermediate};
    \draw( .25,1)node[text depth=0pt]{\rotatebox{90}{\scriptsize Accretion flow}};
    \draw(-.25,1)node[text depth=0pt]{\rotatebox{90}{\scriptsize Chromosphere}};
    \draw(0,-.25)node[text depth=0pt]{\scriptsize\it Transparent Lagrangian interface};
\end{tikzpicture}
\vspace{-.5\baselineskip}
					\caption{"Hybrid" simulation setup and boundary conditions.\label{DESA_ref_HSetup}}
				\end{figure}

				In this Section, the plasma model includes collisional-radiative ionisation (see Section \ref{DESA_ref_ionisation}). The radiation-matter coupling is described within the intermediate regime (see Section \ref{DESA_ref_Zeta}) and the outer radiation flux is set to $c\times E_\mathrm{r}^\mathrm{out}/4$. The goal of this last setup (see Figure \ref{DESA_ref_HSetup}) is to inspect the net effect of the matter-radiation coupling. We have therefore chosen not to consider any chromospheric activity\referee{. Following the preliminary process of the previous setup (see Section \ref{DESA_ref_CL-Setup}), the outer zone is first inactivated, and the radiative hydrostatic equilibrium is computed in the inner zone; once the stationary regime is reached, the accretion process is launched. A key advantage of this process is that nothing is needed to maintain the chromospheric structure, which can therefore freely evolve depending on the physical processes in play only.}

			\subsubsection{Ionisation model}

				We have tested in this setup the effect of the time-dependent ionisation through radiative ionisation/recombination and collisional ionisation with a time-dependent formulation (see Section \ref{DESA_ref_ionisation} for more details).\\

\begin{figure}[htb]
					\centering
					\includegraphics[scale=.8]{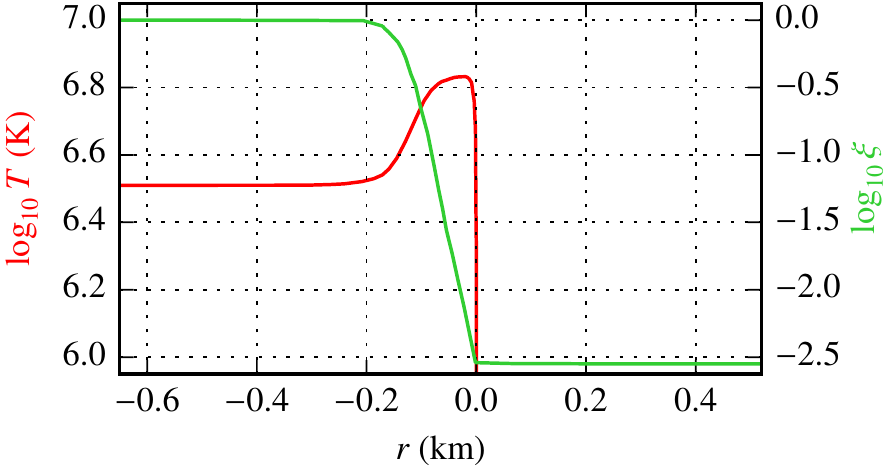}
					\caption{Electron ionisation rate ($\xi=\frac{n_\mathrm{e}}{n_\mathrm{H}+2n_\mathrm{He}+n_\mathrm{M}}$, \texttt{light green}) and temperature (\texttt{red}) profiles zoomed on the reverse shock front in the early QPO cycle.\label{DESA_ref_zoomNe}}
				\end{figure}

				The main difference brought by a time-dependent calculation of the electron density is a tiny ionisation delay behind the reverse shock front, as shown in Figure \ref{DESA_ref_zoomNe}.

				\noindent At the shock front, the kinetic energy is converted into thermal energy, and \emph{then} a part of this thermal energy is used to ionise the post-shock material with a time scale connected to the ionisation rates; the affected gas layer is up to \SI{0.2}{\km} thick, and thus negligible compared to the whole structure (that is at least \SI{e4}{\km} thick, see Table \ref{DESA_ref_sumuptable}). This justifies the use of a time-independent model for ionisation in the previous setups (W--$\Lambda$ and Chr--$\Lambda$). \referee{\citet{Gunther2007} and \citet{Sacco2008} obtain the same conclusion from different approaches.}\\

				However, compared to the Saha-Brown equilibrium calculations, the use of collisional and radiative rates to derive the equilibrium electron density brings differences in the transition between the (almost) neutral medium and the fully ionised plasma. This transition lays between \num{e3.6} and \SI{e4.2}{\K}. However, such temperatures are only reach by the accreted gas during the cooling instability. Its overall effect is hence negligible. The results presented for the Hybrid case (see Figure \ref{DESA_ref_ResH}) are thus based on this collisional-radiative equilibrium calculation of $n_\mathrm{e}$.

			\subsubsection{Radiation and ionisation feedback}

				\begin{sidewaysfigure*}
					\centering
					\includegraphics[scale=.8]{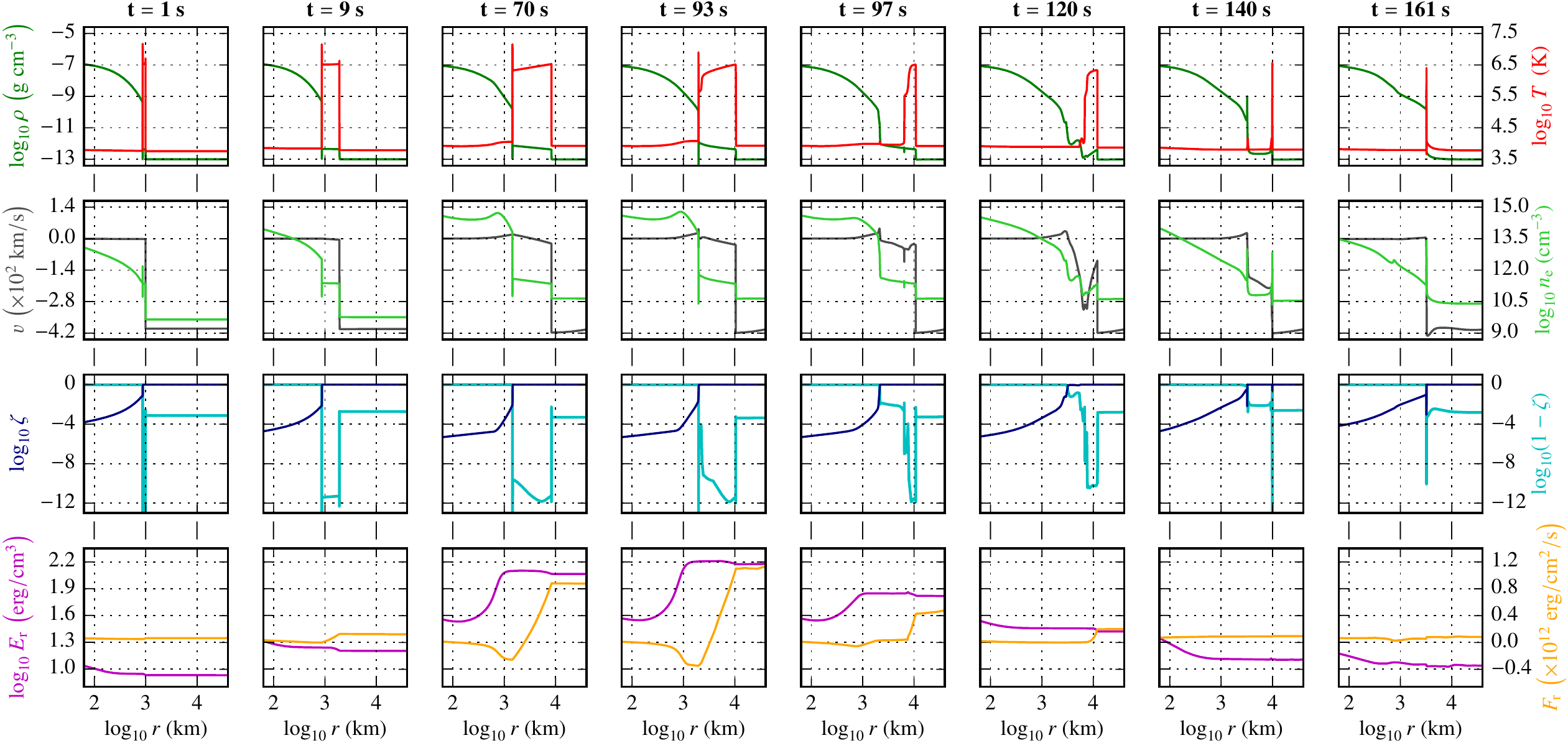}
					\caption{Snapshots of the mass density (\texttt{green}), gas temperature (\texttt{red}), \emph{escape} $\zeta$ (\texttt{dark blue}) and \emph{absorption probabilities} $1-\zeta$ (\texttt{cyan}, see Section \ref{DESA_ref_Zeta}), velocity (\texttt{grey}), electron density (\texttt{light green}), radiation energy density (\texttt{magenta}) and flux (\texttt{orange}) profiles of the first QPO cycle with the "Hybrid" setup. The accreted gas falls from the right to the left on an equilibrium atmosphere.\label{DESA_ref_ResH}}
				\end{sidewaysfigure*}

				The first cycle is presented in Figure \ref{DESA_ref_ResH}; it shows the time variations over \SI{160}{\s} of the gas temperature and mass density, of the photon escape ($\zeta$) and absorption ($1-\zeta$) probabilities and of the radiation energy volumetric density and flux ($E_\mathrm{r}$ and $\vec{F}_\mathrm{\!r}$) (see Section \ref{DESA_ref_Zeta}) for the same snapshots. The next cycles only differ from this first one by the position of the interface between the slab and the chromosphere, as discussed in Section \ref{DESA_ref_chromoheating}.

				The global behaviour follows the trends of the two previous models. However, several effects must be highlighted: a heating of the chromosphere and of the accretion flow, already pointed out by \citet{Calvet1998} and \citet{Costa2017}, and the reduction of the oscillation period and of the post-shock extension. These effects are discussed below.

				\subsubsubsection{QPO cycle reduction}

					Although $1-\zeta$ shows strong variations, its net value beyond the \referee{forward} shock remains negligible, and the post-shock material is in the coronal limit (as in Section \ref{DESA_ref_WL}). The temperature behind the reverse shock is here equal to \SI{3.1e6}{\K}, to be compared to \SI{4e6}{\K} in the reference case. In addition, the compression is enhanced from 4 (W--$\Lambda$ case) to 4.4. As a consequence, the cooling is more efficient: the cooling time is reduced from \SI{400}{\s} down to \SI{220}{\s}, which is compatible with the duration of the cycles. This effect is due to the ionisation/recombination energy cost ($q_\chi$), which is included in the gas energy equation for the Hybrid case, but not for reference case ($q_\chi=0$ in W--$\Lambda$ case, cf. Eq. \eqref{DESA_ref_eqHD} and Section \ref{DESA_ref_ionisation}).

				\subsubsubsection{Radiation energy and flux\label{DESA_ref_ErFr}}

\begin{figure}[htb]
						\centering
						\includegraphics[scale=.8]{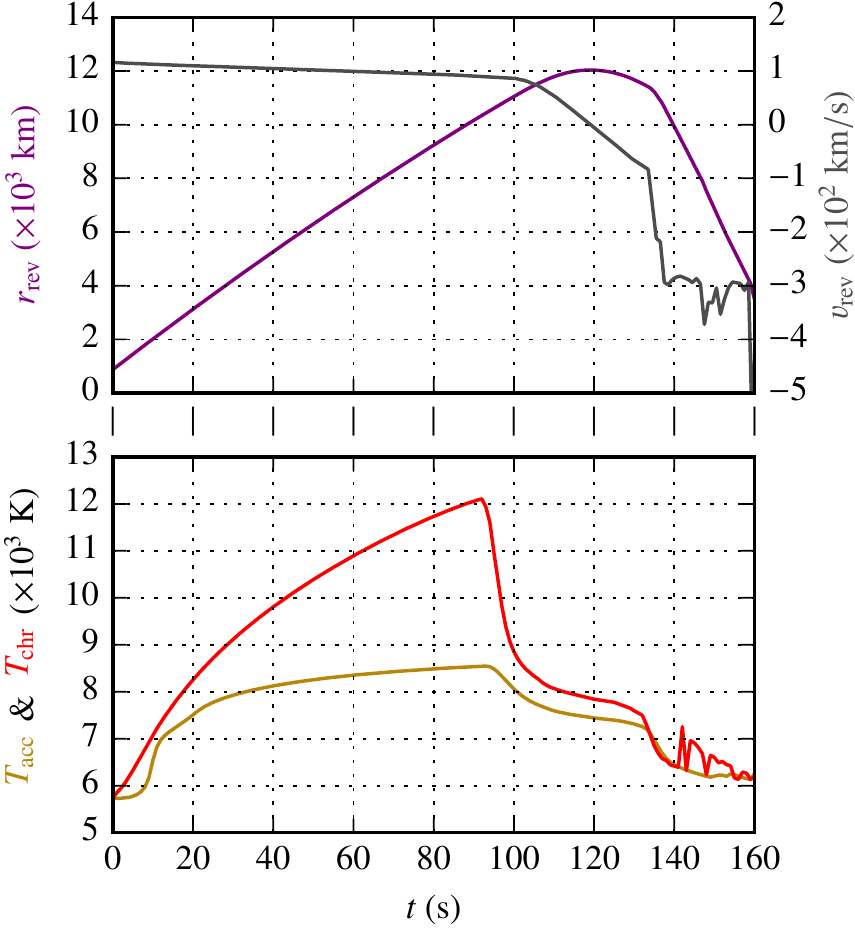}
						\caption{Position of the reverse shock ($r_\mathrm{rev}$, \texttt{purple}) and the corresponding velocity ($v_\mathrm{rev}$, \texttt{grey}) together with the accreted plasma temperature before the hot slab ($T\!_\mathrm{acc}$, \texttt{dark gold}) and the upper chromosphere temperature ($T\!_\mathrm{chr}$, \texttt{red}) for the first QPO cycle in the Hybrid case.\label{DESA_ref_velocities}}
					\end{figure}

					The radiation energy density increases between 9 and \SI{100}{\s}, which corresponds to the growth phase of the hot slab. This increase is however correlated to the upper chromosphere heated up to \SI{12000}{\K} (discussed in Section \ref{DESA_ref_chromobeat} and presented Figure \ref{DESA_ref_velocities}). $E_\mathrm{r}$ remains almost flat in the optically thin post-shock medium, with a value driven by the heated upper chromosphere. In the accretion flow, during the growth of the hot slab, there is a tiny decrease due to the absorption by the accreted material up to $0.5\%$ at \SI{70}{\s} (see Section \ref{DESA_ref_chromoheating}).\\

					\referee{The radiative properties of the inner chromosphere is well described by the diffusive limit: $E_\mathrm{r}\simeq a_\mathrm{R}\,T^4$ and $F_\mathrm{r}\simeq cE_\mathrm{r}/4$.}
					The most peculiar feature of the radiative flux is its linear growth through the post-shock slab. Such a pattern is characteristic of a volume emission by an optically thin medium. As discussed before, the radiative energy in the hot slab is somehow imposed by the heated upper chromosphere: as a consequence, the outgoing radiation flux is $cE_\mathrm{r}/4 \simeq c\,a_\mathrm{R}\,T^4_\mathrm{\!chr}/4$, with $T_\mathrm{\!chr}$ the temperature of the upper chromosphere (cf. Figure \ref{DESA_ref_velocities}). Then, the radiation flux propagates through the accretion flow with negligible changes.\\
					Since $F_\mathrm{\!r}$ is counted negatively towards the star, the flux emitted by the slab is offset by the one produced by the chromosphere. the net radiation flux rises then back in the chromosphere.\\
					\referee{However, the variations of $F_\mathrm{\!r}$ within the slab comes from the interweaving of several radiation sources (the chromosphere, the slab itself and the accretion flow): due to the limitations of the M1 radiation transfer (cf. Section \ref{DESA_ref_RT_eqs}), these variations must be interpreted with care.}

				\subsubsubsection{Chromospheric heating and beating\label{DESA_ref_chromobeat}}

					\begin{figure}[htb]
						\centering
						\includegraphics[scale=.8]{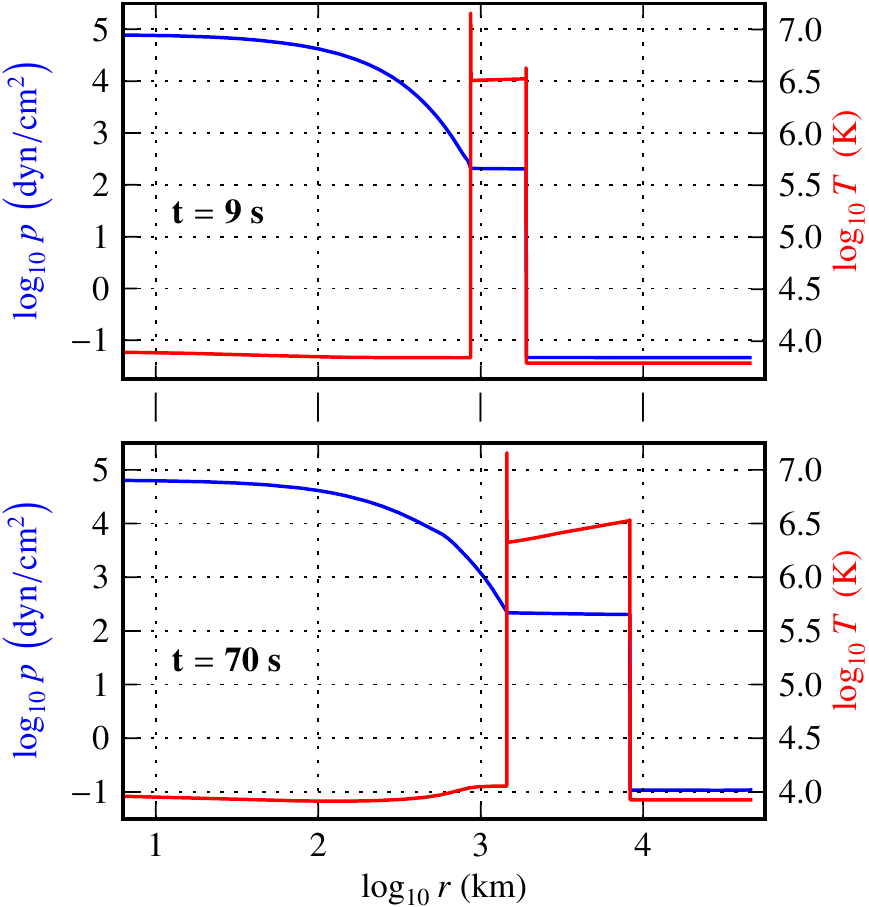}
						\caption{Snapshots of the temperature (\texttt{red}) and pressure (\texttt{blue}) at $9$ (\texttt{top}) and \SI{70}{\s} (\texttt{bottom}) for the Hybrid case.\label{DESA_ref_Heating}}
					\end{figure}

					The upper chromosphere is heated by the radiating post-shock plasma up to \SI{12000}{\K} (Figure \ref{DESA_ref_velocities}). For instance, between $9$ and \SI{70}{\s}, its temperature varies from $7000$ to \SI{10800}{\K} at \SI{800}{\km}, and the pressure increases from $800$ to \SI{2600}{dyn\per\square\cm} at this location (Figure \ref{DESA_ref_Heating}). As a consequence, the whole post-shock structure is pushed upwards from \SI{875}{\km} to \SI{3150}{\km}, thus out of the unperturbed chromosphere \citep[by about \SI{2000}{\km}, see e.g.][]{Vernazza1973}.

					At the end of the cycle, the chromosphere is not heated any more and the slab buries back into the atmosphere. The expected behaviour is an oscillation of the slab \emph{burial} with the same periodicity as QPOs, since it originates from the hot post-shock plasma radiation. At the end of this first cycle, the chromosphere does not recover its initial thickness: this effect does not affect the post-shock dynamics and cycle characteristics. All these effects are overestimated in a 1D model. However, this study shows that the general question of the (un)burial, which is important for X-ray observations, can only be addressed within a model that takes into account the radiative heating of the chromosphere by the hot slab.

				\subsubsubsection{Accretion flow pre-heating\label{DESA_ref_chromoheating}}

					While reaching the hydro-radiative steady state of the chromosphere, the flow has been homogeneously heated from \SI{3000}{\K} to \SI{5730}{\K} before the start of the accretion process. During the cycle, the accretion leads to an additive heating of the flow up to $\sim\SI{8500}{\K}$ (at $t=\SI{93}{\s}$). These effects are quantified in Figure \ref{DESA_ref_velocities}, which reports the time variations of the position and velocity of the interface between the hot slab and the accretion flow, as well as the temperatures of the heated chromosphere and of the pre-shock material. \referee{The use of the escape probability formalism (see Section \ref{DESA_ref_Zeta}) induces a dependence of the absorption by the accretion flow with the section of the column; changing this section from \SI{1000}{\km} to \SI{10000}{\km} for instance will vary the parameter $\zeta$ from $1-\num{5e-3}$ to $1-\num{5e-2}$, increasing the absorption and thus the radiative heating of the pre-shock flow.}

					Such preheating has already been pointed out by other authors \citep{Calvet1998,Costa2017}. In these works, this heating is induced by radiation coming from the hot slab through photo-ionisation. Although radiative cooling of the accretion flow may be included in some cases, the radiation transfer is not taken into account. Depending on the conditions, the pre-shock temperature may reach from \SI{20000}{\K} (in CG98) up to \SI{e5}{\K} (in Co17) close to the reverse shock (up to \SI{e4}{\km}). In the latter, this precursor is preceded by a flatter and cooler ($\sim\SI{e4}{\K}$) zone with an extension of \SI{e5}{\km}, thus smaller than ours ($>\SI{e5}{\km}$).

					Our simulation shows that part of the heating is a consequence of the chromospheric radiation already in play before the start of the accretion. The analysis of the variation of the radiative energy indicates that an additional heating operates during the development of the hot slab. However, as we do not include any dependence with the wavelength, it remains very difficult to discriminate in details the role played by the radiation emitted by the hot slab (X-rays) and from the (heated) chromosphere (UV-visible). Complementary information will be given by the synthetic spectra computed as a post-process of the hydrodynamics structures (Section \ref{DESA_ref_spectra}).

			 \subsubsubsection{X-ray luminosity}

					The X-ray luminosity of the system is computed following the method described in Section \ref{DESA_ref_XRL}. Its time variation (in unit of $\tau_\mathrm{cycle} = \SI{160}{\s}$) is reported in Figure \ref{DESA_ref_Flux_Lambda} for comparison with the two previous cases. Compared to the reference case, in addition with a shortening of the period, this case presents a more pronounced radiative collapse (\SIrange{70}{90}{\s}), followed by a chaotic collapse (\SIrange{90}{160}{\s}). The time average of the radiative surface luminosity is here equal to \SI{1.2e9}{\erg\per\square\cm\per\s}, which represents $30\%$ of the mechanical energy flux (cf. Figure \ref{DESA_ref_Flux_Lambda}).

			\subsubsection{SYNSPEC monochromatic emergent intensity\label{DESA_ref_spectra}}

				As this simulation is performed using only one group of radiation frequencies, it is interesting to analyse more precisely the details of the previous radiative heating via its feedback on the monochromatic emergent intensity.

				To this purpose, the hydrodynamic structures has been post-processed with the SYNSPEC code (Section \ref{DESA_ref_SYNSPEC}). For consistency purpose, we take the atomic data already used for the calculation of the average opacities (see Section \ref{DESA_ref_radsources} and Appendix \ref{DESA_ref_app_opacities}). We thus estimate the specific intensity $I_{\!\lambda}^\shortparallel$ (in \si{\erg\per\square\cm\per\s\per\AA\per\sr}) along the direction of the column. Since the line profile behaviour is not investigated here, velocity effects are neglected.

				It is important to recall that a quantitative comparison of this synthetic spectrum with observations, especially in the X-rays \citep[see e.g.][]{Gudel2007a,Robrade2007,Drake2009} would require NLTE and 3D radiative transfer post-processing. Nonetheless, using 1D radiative transfer and the LTE approximation is here interesting as it corroborates or not the general accepted trends, e.g. a strong X-ray emission and an excess of luminosity in the UV-VIS range \citep{Calvet1998,Brickhouse2010,Ingleby2013}.

				\begin{figure}[htb]
					\centering
					\includegraphics[scale=.8]{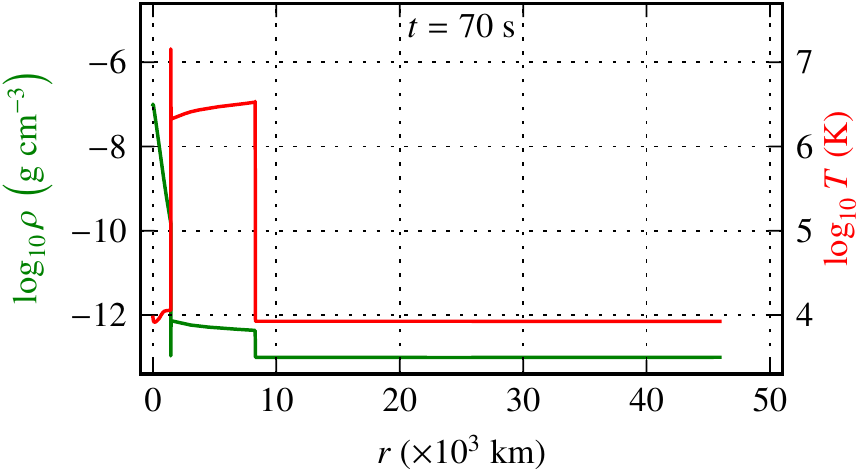}
					\caption{Snapshot of the density (\texttt{green}) and temperature (\texttt{red}) profiles at $t=\SI{70}{\s}$ with the "Hybrid" setup, post-processed hereafter.\label{DESA_ref_snapshot_H70}}
				\end{figure}

				\begin{figure}[htb]
					\centering
					\includegraphics[scale=.8]{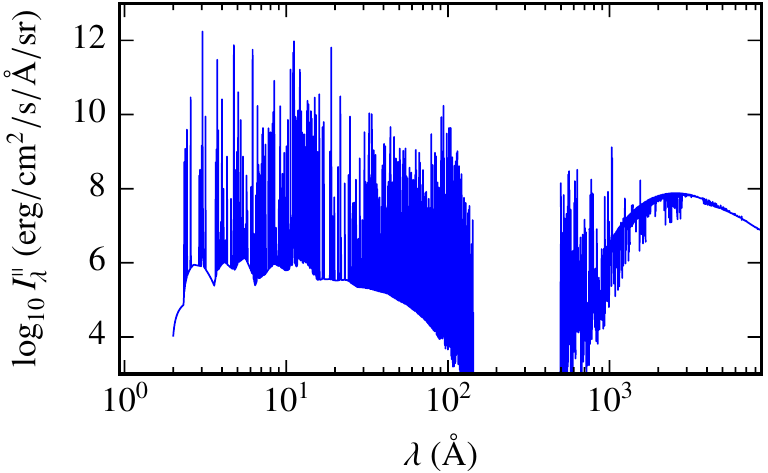}
					\caption{Specific intensity $I_{\!\lambda}^\shortparallel$ parallel to the column during the QPO cycle of the \emph{Hybrid} model (at $t=\SI{70}{\s}$, see Figure \ref{DESA_ref_snapshot_H70}).\label{DESA_ref_spectrum_H70}}
				\end{figure}

				A typical spectrum emerging from \SI{4.6e4}{\km} (located within the accretion flow) is reported in Figure \ref{DESA_ref_spectrum_H70}). It is computed from a snapshot ($t=\SI{70}{\s}$) of the \emph{Hybrid} model (see Figures \ref{DESA_ref_ResH} and \ref{DESA_ref_snapshot_H70}). At this stage, the chromosphere extends up to \SI{1.4e3}{\km}, the hot plasma from \SI{1.4e3}{\km} to \SI{8.3e3}{\km} and the accretion flow from \SI{8.3e3}{\km} to \SI{1e5}{\km}. The intensity that emerges from this layer presents three characteristic spectral bands:
				\begin{itemize}[nosep]
					\item in the range \SIrange{1}{100}{\AA} (X-rays), the bump is attributed to the hot post-shock plasma, with intense lines up to \SI{e12}{\erg\per\square\cm\per\s\per\AA};
					\item in the range \SIrange{100}{900}{\AA} (EUV), radiation is efficiently absorbed by the inflow;
					\item in the range \SIrange{900}{10000}{\AA} (UV+Vis+IR), the second bump is attributed to the \emph{heated} stellar chromosphere and photosphere, i.e. a black body at $T\simeq\SI{11300}{\K}$ (cf. Figure \ref{DESA_ref_velocities}).
				\end{itemize}

				\begin{figure}[htb]
					\centering
					\includegraphics[scale=.8]{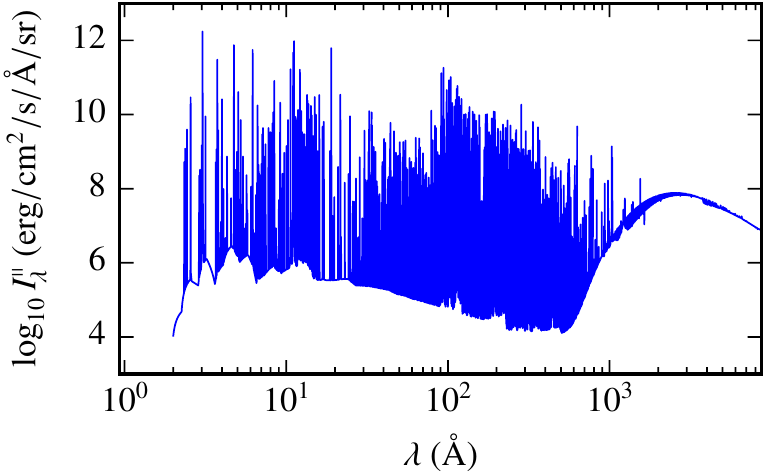}
					\caption{Specific intensity $I_{\!\lambda}^\shortparallel$ emerging from the reverse shock front during the QPO cycle of the \emph{Hybrid} model (at $t=\SI{70}{\s}$, see Figure \ref{DESA_ref_snapshot_H70}).\label{DESA_ref_spectrum_H70cut}}
				\end{figure}

				The strong absorption of the EUV radiation is due to the huge optical depth of the accretion flow\footnote{at $\rho=\SI{e-13}{\gram\per\cubic\cm}$ and $T\simeq\SIrange{5000}{8000}{\K}$.}. This effect may then be attenuated in the case of a bent column or when the observation is performed side-on and not along the column. This absorption effect on the spectrum is illustrated in Figure \ref{DESA_ref_spectrum_H70cut}, which presents the intensity emerging \emph{right after the reverse shock front}, at $r=\SI{8.3e3}{\km}$. This figure shows that this absorption also affects, to a lesser degree, the visible spectrum originating from the chromosphere. This must be considered when interpreting the UV excess \citep[see e.g.][]{Calvet1998,Hartmann2016,Colombo2019}. Note that a pre-heating of the accretion flow is expected as a result of the EUV absorption. A pre-heating is also obtained independently by AstroLabE (Section \ref{DESA_ref_chromoheating}); however, a one-to-one correspondence would require a multi-group description of the radiation field in AstroLabE.

				\begin{figure}[htb]
					\centering
					\includegraphics[scale=.8]{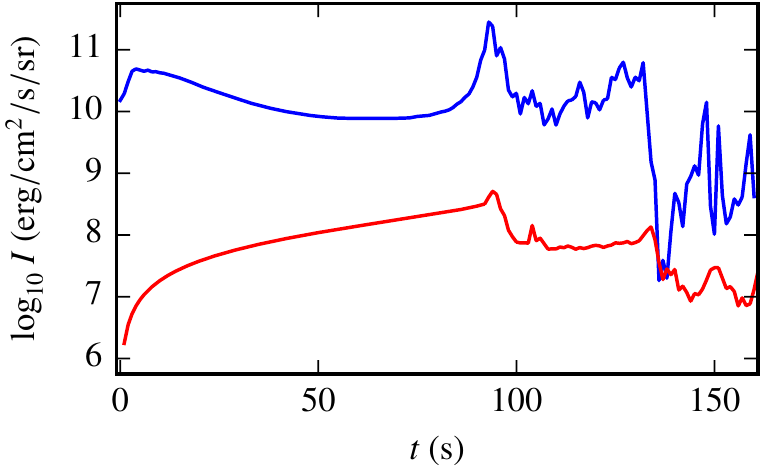}
					\caption{Time-variation of the \SIrange{2}{27}{\AA} (LTE) integrated X-ray outgoing intensity ($I_\mathrm{\!X}^\shortparallel$, \texttt{blue}) and of the optically thin post-shock emission ($I_\mathrm{\!\Lambda}$ \texttt{red}) during a QPO cycle for the \emph{Hybrid} setup.\label{DESA_ref_time_Hflux}}
				\end{figure}

				We compute from $I_{\!\lambda}^\shortparallel$ (Figure \ref{DESA_ref_snapshot_H70}) the net X-ray outgoing intensity ($I_\mathrm{\!X}^\shortparallel$) and the corresponding coronal quantity ($I_\mathrm{\!\Lambda}$):
				\begin{equation}
					I_\mathrm{\!X}^\shortparallel = \int_2^{27}\hspace{-1em}I_{\!\lambda}^\shortparallel\,d_{\!}\lambda
					\qquad\text{and}\qquad
					I_\mathrm{\!\Lambda} = \frac{\mathcal{L}_\mathrm{\!\Lambda}}{4\pi}
				\end{equation}
				The time variations of these two quantities, reported in Figure \ref{DESA_ref_time_Hflux}, present similar characteristics. However, the values derived by SYNSPEC are higher by about two to three orders of magnitude. This discrepancy is either imputable to the LTE approximation or to the assumed 1D plane-parallel geometry. Thus our synthetic spectra can't be used for quantitative comparison with observations.

	\section{Refining the models\label{DESA_ref_caveats}}

		\subsection{A more realistic chromosphere}

			It should be pointed out that this study uses a solar model for the chromosphere with acoustic heating. Compared to the description of this heating, a more important improvement would be to consider a realistic T Tauri chromospheric model, which is today not very well known. This may affect the ionisation (and then gas pressure with another chemical abundances) as well as slab characteristics (through gravity) and radiation effects (through opacities and incoming luminosity). Our results are then to be considered qualitatively and not quantitatively.

		\subsection{Improvements of the radiation model}

			We use in this work radiation momenta equations with the M1 closure relation. Although this is already a strong improvement compared to other approaches like the diffusion model, it could be improved by using radiation half-fluxes (i.e. the inward an outward components of the radiation flux). This should disentangle the radiation flux coming from the star and from the post-shock structure.\\
			The M1 closure relation allows the radiation field to reach at most one direction of anisotropy; half-fluxes can extend it to two, i.e. the maximum number of anisotropy directions reachable in 1D. Half-fluxes (along with M1) would then be equivalent to the momenta equations with the M2 closure relation \citep{Feugeas2004}, without its prohibitive numerical cost.\\
			The M1 model and its limits have been thoroughly studied \citep[see e.g.][]{Levermore1996,Dubroca1999,Feugeas2004}. The behaviour of this model \emph{along with} half-fluxes needs however to be examined.

			More important is the approximation made with the \emph{monogroup} approach used in this work. The whole spectrum is then approximated as a black body providing the adequate opacity averages. However, our computed spectra emerging from accretion structures are expected to present three discernible frequency groups:
			\begin{itemize}[nosep]
				\item up to the visible domain, the spectrum is dominated by the black body emerging from the stellar photosphere;
				\item the EUV band is expected to be depleted due to high absorption by the accreted gas;
				\item the X-ray band is thought to be optically thin and to have the \referee{hot slab} signature on it.
			\end{itemize}
			Although the \emph{multi-group} approach is numerically heavier, it will improve the study of the consequences of the radiation absorption by the surrounding medium. A consequence of the X-ray and EUV absorption by the cold accretion flow is the presence of a radiative precursor. Such a phenomenon cannot be obtained through a monogroup approach. Moreover, a 3 groups approach will provide a better description of the feedback of the hot slab on the stellar chromosphere.

		\subsection{NLTE effects in radiation hydrodynamics and in synthetic spectra}

			Two other points may be improved. First, the transition model ($\zeta$) remains qualitative and may need to be extended to the ionisation calculation. The work done by \citet{Carlsson2012} offers paths to reach such consistency and may need to be investigated further. A better model of both the LTE transfer, line cooling and intermediate regimes may demand dedicated NLTE opacities, namely plasma \emph{emissivity} (equivalent to $n_e\,n_\mathrm{H}\,\Lambda$), radiation energy absorption ($\kappa_\mathrm{P}$) and radiation flux sinking ($\kappa_\mathrm{R}$). Moreover, all these quantities, computed with a radiative-collisionnal model, have to be averaged over adequate weighting functions. Due to recent progresses in this topic \citep{Rodriguez2018}, new results are expected in a near future. Independently, a NLTE description should be used to compute the emerging spectra: this work is already in progress using TLUSTY code.

	\section{Conclusion}

		In this study, we used 1D simulations with detailed physics to check the validity of the two following common assumptions in accretion shocks simulations: the stellar atmosphere can be either modelled by a hydrostatic or a steady hydrodynamic structure, and the dynamics of accretion shocks is governed by optically thin radiation transfer. We checked first that we are able to recover previous results \citep[W--$\Lambda$ case, Section \ref{DESA_ref_WL}]{Sacco2008} and tested independently each of these assumptions (Chr--$\Lambda$ case, Section \ref{DESA_ref_CL}, and Hybrid case, Section \ref{DESA_ref_Rad}). Each of them proves to have a non-negligible impact on the typical characteristics of the accretion dynamics and on the estimation of its X-ray surface luminosity. This one varies between $1$ and \SI{4e9}{\erg\per\square\cm\per\s}. Taking as a reference the radiative power of \citet{Brickhouse2010}, we derive a section of the accretion spot from \num{3e20} to \SI{1e21}{\square\cm}, corresponding to a filling factor of the solar disk between $2$ and $8\%$\referee{, i.e. a stream composed of $\sim\num{e4}$ fibrils of radius $L_\mathrm{c}$ (cf. Section \ref{DESA_ref_strategy}) or a column of radius $100\times L_\mathrm{c}$, supposing that the global dynamics of the system is not influenced by this larger section of the column through radiative effects}.

		In the case of the chromosphere which is heated by acoustic perturbations that degenerate into small shock waves, we have shown that these perturbations do not strongly modify the cycle period compared to the reference case. However, the cycle becomes chaotic due to the generation of secondary shock waves. As a result, the relative duration of the hot phase in the cycle remains longer, and thus the variability in the X-rays is less pronounced than for the reference case. To be detected, it would require a better sensitivity of the photometric measurements.\\

		In the case of an initially steady atmosphere at radiative equilibrium, the coupling between the radiation and the hydrodynamics leads to:
		\begin{itemize}[nosep]
			\item a radiative feedback (heating) of the atmosphere, which successively expands and retracts, inducing in particular an unburial of the column, which is favorable to the lateral escape of the X-ray emitted from the hot slab;
			\item a chaotic radiative collapse, with an impact on the time variation of the X-ray flux (Figure \ref{DESA_ref_Flux_Lambda});
			\item a radiative pre-heating of the incoming flow, over the length of the simulation box.
		\end{itemize}
		Moreover the inclusion of ionization in the energy balance leads to important effects in the post-shock temperature that modify the cooling efficiency and therefore the cycle duration.

		In this hybrid case, we computed at LTE the radiative intensity emerging from the location of the reverse shock (Figure \ref{DESA_ref_spectrum_H70}) as also from the outer boundary (Figure \ref{DESA_ref_spectrum_H70cut}). The flux is characterized by --1-- a huge number of atomic lines in the X rays, --2-- a near blackbody profile in the Visible, with the presence of emission and absorption lines, --3-- a EUV component which is very strong at the position of the reverse shock and disappears at the outer boundary, due to the importance of the absorption.

		This study could be completed with a more complete simulation that would include both a dynamically-heated chromosphere and the hybrid setup. However, it appears at this stage more important to take into account a NLTE radiative description based on adapted opacities and radiative power losses. Another necessary improvement will be through a multi-group radiation transfer to catch at least the effect of EUV absorption and X-ray radiative losses on the structure of the column, and to analyse the possibility of a radiative precursor which could pre-heat the incoming flow.
The study is also to be extended to multi-dimensional simulations in order to check the effects of both radiation and magnetic field closer to the real picture \citep{Orlando2010,Orlando2013,Matsakos2013,Matsakos2014b}.

		\begin{acknowledgements}
			We express our gratitude to Jason Ferguson for providing us with the molecular LTE opacity tables used in this work\referee{ and to Franck Delahaye for his contribution to the atomic LTE ionisation data used for both the opacity tables and for the radiative transfer post processing}. We thank Rafael Rodriguez for sharing with us helpful preliminary results about NLTE microscopic collisional-radiative data, Salvatore Orlando for reading the manuscript, Ziane Izri for its contribution at the beginning of this project and Christophe Sauty for helpful discussions.\\
			I.H. thanks the Physics Department of Sorbonne Université for his visiting professorship.\\
			This work was supported by the french ANR StarShock and LabEx Plas@Par projects (resp. ANR--08--BLAN--0263--07 and ANR--11--IDEX--0004--02), PICS 6838, Programme National de Physique Stellaire of CNRS/INSU and Observatoire de Paris.
		\end{acknowledgements}

	\newcommand*\astl{AstL}
	\newcommand*\rvma{RvMA}
	\bibliographystyle{aa}\bibliography{DE_SA}

\begin{thebibliography}{100}
\expandafter\ifx\csname natexlab\endcsname\relax\def\natexlab#1{#1}\fi

\bibitem[{Alfv{\'e}n \& Lindblad(1947)}]{Alfven1947}
Alfv{\'e}n, H. \& Lindblad, B. 1947, \mnras, 107, 211

\bibitem[{Argiroffi {et~al.}(2007)Argiroffi, Maggio, \& Peres}]{Argiroffi2007}
Argiroffi, C., Maggio, A., \& Peres, G. 2007, \aap, 465, L5

\bibitem[{Argiroffi {et~al.}(2009)Argiroffi, Maggio, Peres, Drake,
  L{\'o}pez-Santiago, Sciortino, \& Stelzer}]{Argiroffi2009}
Argiroffi, C., Maggio, A., Peres, G., {et~al.} 2009, \aap, 507, 939

\bibitem[{Auer(2003)}]{Auer2003}
Auer, L.~H. 2003, in Stellar Atmosphere Modeling, Vol. 288 (ASPC), 3--15

\bibitem[{Ayres(1979)}]{Ayres1979}
Ayres, T.~R. 1979, \apj, 228, 509

\bibitem[{Batalha \& Basri(1993)}]{Batalha1993}
Batalha, C.~C. \& Basri, G. 1993, \apj, 412, 363

\bibitem[{Biermann(1946)}]{Biermann1946}
Biermann, L. 1946, Naturwissenschaften, 33, 118

\bibitem[{Bonito {et~al.}(2014)Bonito, Orlando, Argiroffi, Miceli, Peres,
  Matsakos, Stehl{\'e}, \& Ibgui}]{Bonito2014b}
Bonito, R., Orlando, S., Argiroffi, C., {et~al.} 2014, \apj, 795, L34

\bibitem[{Bouquet {et~al.}(2004)Bouquet, Stehl{\'e}, Koenig, Chi{\`e}ze,
  Benuzzi-Mounaix, Batani, Leygnac, Fleury, Merdji, Michaut, Thais, Grandjouan,
  Hall, Henry, Malka, \& Lafon}]{Bouquet2004}
Bouquet, S., Stehl{\'e}, C., Koenig, M., {et~al.} 2004, Phys. Rev. Lett., 92,
  225001

\bibitem[{Bouvier {et~al.}(1995)Bouvier, Covino, Kovo, Martin, Matthews,
  Terranegra, \& Beck}]{Bouvier1995}
Bouvier, J., Covino, E., Kovo, O., {et~al.} 1995, \aap, 299, 89

\bibitem[{Brickhouse {et~al.}(2010)Brickhouse, Cranmer, Dupree, Luna, \&
  Wolk}]{Brickhouse2010}
Brickhouse, N.~S., Cranmer, S.~R., Dupree, A.~K., Luna, G. J.~M., \& Wolk,
  S.~J. 2010, \apj, 710, 1835

\bibitem[{Brown(1973)}]{Brown1973}
Brown, J.~C. 1973, \solphys, 29, 421

\bibitem[{Calvet(1983)}]{Calvet1983}
Calvet, N. 1983, \rmxaa, 7, 169

\bibitem[{Calvet {et~al.}(1984)Calvet, Basri, \& Kuhi}]{Calvet1984}
Calvet, N., Basri, G., \& Kuhi, L.~V. 1984, \apj, 277, 725

\bibitem[{Calvet \& Gullbring(1998)}]{Calvet1998}
Calvet, N. \& Gullbring, E. 1998, \apj, 509, 802

\bibitem[{Carlsson \& Leenaarts(2012)}]{Carlsson2012}
Carlsson, M. \& Leenaarts, J. 2012, \aap, 539, A39

\bibitem[{Charignon \& Chi{\`e}ze(2013)}]{Charignon2013}
Charignon, C. \& Chi{\`e}ze, J.-P. 2013, \aap, 550, A105

\bibitem[{Chevalier \& Imamura(1982)}]{Chevalier1982}
Chevalier, R.~A. \& Imamura, J.~N. 1982, \apj, 261, 543

\bibitem[{Chi{\`e}ze {et~al.}(2012)Chi{\`e}ze, de~S{\'a}, \&
  Stehl{\'e}}]{Chieze2012}
Chi{\`e}ze, J.-P., de~S{\'a}, L., \& Stehl{\'e}, C. 2012, EAS Publications
  Series, 58, 143

\bibitem[{Colombo {et~al.}(2019)Colombo, Ibgui, Orlando, Rodriguez, Espinosa,
  Gonz{\'a}lez, Stehlé, de~S{\'a}, Argiroffi, Bonito, \& Peres}]{Colombo2019}
Colombo, S., Ibgui, L., Orlando, S., {et~al.} 2019, \aap, submitted

\bibitem[{Colombo {et~al.}(2016)Colombo, Orlando, Peres, Argiroffi, \&
  Reale}]{Colombo2016}
Colombo, S., Orlando, S., Peres, G., Argiroffi, C., \& Reale, F. 2016, \aap,
  594, A93

\bibitem[{Costa {et~al.}(2017)Costa, Orlando, Peres, Argiroffi, \&
  Bonito}]{Costa2017}
Costa, G., Orlando, S., Peres, G., Argiroffi, C., \& Bonito, R. 2017, \aap,
  597, A1

\bibitem[{Cram(1979)}]{Cram1979}
Cram, L.~E. 1979, \apj, 234, 949

\bibitem[{Curran {et~al.}(2011)Curran, Argiroffi, Sacco, Orlando, Peres, Reale,
  \& Maggio}]{Curran2011}
Curran, R.~L., Argiroffi, C., Sacco, G.~G., {et~al.} 2011, \aap, 526, A104

\bibitem[{de~S\'a(2014)}]{deSaPhD2014}
de~S\'a, L. 2014, PhD thesis, Universit\'e de Paris VI

\bibitem[{{de S{\'a}} {et~al.}(2012){de S{\'a}}, {Chi{\`e}ze}, {Stehl{\'e}},
  {Hubeny}, {Delahaye}, \& {Lanz}}]{deSa2012}
{de S{\'a}}, L., {Chi{\`e}ze}, J.-P., {Stehl{\'e}}, C., {et~al.} 2012, in
  SF2A-2012: Proceedings of the Annual meeting of the French Society of
  Astronomy and Astrophysics, ed. S.~{Boissier}, P.~{de Laverny},
  N.~{Nardetto}, R.~{Samadi}, D.~{Valls-Gabaud}, \& H.~{Wozniak}, 309--312

\bibitem[{de~S\'a {et~al.}(2014)de~S\'a, Chi\`eze, Stehl\'e, Matsakos, Ibgui,
  Lanz, \& Hubeny}]{deSa2014}
de~S\'a, L., Chi\`eze, J.-P., Stehl\'e, C., {et~al.} 2014, in European Physical
  Journal Web of Conferences, Vol.~64, 04002

\bibitem[{Dorfi \& Drury(1987)}]{Dorfi1987}
Dorfi, E.~A. \& Drury, L.~O. 1987, JCoPh, 69, 175

\bibitem[{Drake(2005)}]{Drake2005a}
Drake, J.~J. 2005, in 13th Cambridge Workshop on Cool Stars, Stellar Systems
  and the Sun, 519

\bibitem[{Drake {et~al.}(2009)Drake, Ratzlaff, Laming, \& Raymond}]{Drake2009}
Drake, J.~J., Ratzlaff, P.~W., Laming, J.~M., \& Raymond, J.~C. 2009, \apj,
  703, 1224

\bibitem[{Dubroca \& Feugeas(1999)}]{Dubroca1999}
Dubroca, B. \& Feugeas, J.-L. 1999, CRAS Paris S{\'e}rie 1, 329, 915

\bibitem[{Dumont {et~al.}(1973)Dumont, Heidmann, Kuhi, \& Thomas}]{Dumont1973}
Dumont, S., Heidmann, N., Kuhi, L.~V., \& Thomas, R.~N. 1973, \aap, 29, 199

\bibitem[{Feigelson \& Montmerle(1999)}]{Feigelson1999}
Feigelson, E.~D. \& Montmerle, T. 1999, \araa, 37, 363

\bibitem[{Ferguson {et~al.}(2005)Ferguson, Alexander, Allard, Barman, Bodnarik,
  Hauschildt, Heffner-Wong, \& Tamanai}]{Ferguson2005}
Ferguson, J.~W., Alexander, D.~R., Allard, F., {et~al.} 2005, \apj, 623, 585

\bibitem[{Feugeas(2004)}]{Feugeas2004}
Feugeas, J.-L. 2004, LPB, 22, 121

\bibitem[{Fitzpatrick(1996)}]{Fitzpatrick1996}
Fitzpatrick, E.~L. 1996, \apj, 473, L55

\bibitem[{Fritsch \& Butland(1984)}]{Fritsch1984}
Fritsch, F.~N. \& Butland, J. 1984, SIAM J. Sci. and Stat. Comput., 5, 300

\bibitem[{Grevesse \& Sauval(1998)}]{Grevesse1998}
Grevesse, N. \& Sauval, A.~J. 1998, \ssr, 85, 161

\bibitem[{G{\"u}del {et~al.}(2007)G{\"u}del, Skinner, Mel'nikov, Audard,
  Telleschi, \& Briggs}]{Gudel2007a}
G{\"u}del, M., Skinner, S.~L., Mel'nikov, S.~Y., {et~al.} 2007, \aap, 468, 529

\bibitem[{Gullbring {et~al.}(2000)Gullbring, Calvet, Muzerolle, \&
  Hartmann}]{Gullbring2000}
Gullbring, E., Calvet, N., Muzerolle, J., \& Hartmann, L.~W. 2000, \apj, 544,
  927

\bibitem[{G{\"u}nther {et~al.}(2010)G{\"u}nther, Lewandowska, Hundertmark,
  Steinle, Schmitt, Buckley, Crawford, O'Donoghue, \& Vaisanen}]{Gunther2010}
G{\"u}nther, H.~M., Lewandowska, N., Hundertmark, M. P.~G., {et~al.} 2010,
  \aap, 518, A54

\bibitem[{G{\"u}nther {et~al.}(2006)G{\"u}nther, Liefke, Schmitt, Robrade, \&
  Ness}]{Gunther2006}
G{\"u}nther, H.~M., Liefke, C., Schmitt, J. H. M.~M., Robrade, J., \& Ness,
  J.~U. 2006, \aap, 459, L29

\bibitem[{G{\"u}nther {et~al.}(2007)G{\"u}nther, Schmitt, Robrade, \&
  Liefke}]{Gunther2007}
G{\"u}nther, H.~M., Schmitt, J. H. M.~M., Robrade, J., \& Liefke, C. 2007,
  \aap, 466, 1111

\bibitem[{Hartmann {et~al.}(2016)Hartmann, Herczeg, \& Calvet}]{Hartmann2016}
Hartmann, L.~W., Herczeg, G.~J., \& Calvet, N. 2016, \araa, 54, 135

\bibitem[{Hubeny \& Lanz(1995)}]{Hubeny1995}
Hubeny, I. \& Lanz, T. 1995, \apj, 439, 875

\bibitem[{Hubeny \& Lanz(2017)}]{Hubeny2017}
Hubeny, I. \& Lanz, T. 2017, eprint arXiv:1706.01859,

\bibitem[{Hubeny \& Mihalas(2014)}]{Hubeny2014}
Hubeny, I. \& Mihalas, D. 2014, {Theory of Stellar Atmospheres An Introduction
  to Astrophysical Non-equilibrium Quantitative Spectroscopic Analysis}, 1st
  edn. (Princeton University Press)

\bibitem[{Huenemoerder {et~al.}(2007)Huenemoerder, Kastner, Testa, Schulz, \&
  Weintraub}]{Huenemoerder2007}
Huenemoerder, D.~P., Kastner, J.~H., Testa, P., Schulz, N.~S., \& Weintraub,
  D.~A. 2007, \apj, 671, 592

\bibitem[{Hui \& Gnedin(1997)}]{Hui1997}
Hui, L. \& Gnedin, N.~Y. 1997, \mnras, 292, 27

\bibitem[{Ibgui {et~al.}(2013)Ibgui, Hubeny, Lanz, \& Stehl{\'e}}]{Ibgui2013}
Ibgui, L., Hubeny, I., Lanz, T., \& Stehl{\'e}, C. 2013, \aap, 549, A126

\bibitem[{Ingleby {et~al.}(2013)Ingleby, Calvet, Herczeg, Blaty, Walter,
  Ardila, Alexander, Edwards, Espaillat, Gregory, Hillenbrand, \&
  Brown}]{Ingleby2013}
Ingleby, L., Calvet, N., Herczeg, G.~J., {et~al.} 2013, \apj, 767, 112

\bibitem[{Jess {et~al.}(2015)Jess, Morton, Verth, Fedun, Grant, \&
  Giagkiozis}]{Jess2015}
Jess, D.~B., Morton, R.~J., Verth, G., {et~al.} 2015, \ssr, 190, 103

\bibitem[{Jiang {et~al.}(2014{\natexlab{a}})Jiang, Stone, \&
  Davis}]{Jiang2014b}
Jiang, Y.-F., Stone, J.~M., \& Davis, S.~W. 2014{\natexlab{a}}, \apj, 796, 106

\bibitem[{Jiang {et~al.}(2014{\natexlab{b}})Jiang, Stone, \&
  Davis}]{Jiang2014a}
Jiang, Y.-F., Stone, J.~M., \& Davis, S.~W. 2014{\natexlab{b}}, \apjs, 213, 7

\bibitem[{Johns-Krull(2007)}]{Johns-Krull2007}
Johns-Krull, C.~M. 2007, \apj, 664, 975

\bibitem[{Johns-Krull {et~al.}(1999)Johns-Krull, Valenti, Hatzes, \&
  Kanaan}]{Johns-Krull1999}
Johns-Krull, C.~M., Valenti, J.~A., Hatzes, A.~P., \& Kanaan, A. 1999, \apj,
  510, L41

\bibitem[{Judge(2006)}]{Judge2006}
Judge, P. 2006, in Solar MHD Theory and Observations: A High Spatial Resolution
  Perspective, Vol. 354 (Astronomical Society of the Pacific Conference
  Series), 259

\bibitem[{Kalkofen(2007)}]{Kalkofen2007}
Kalkofen, W. 2007, \apj, 671, 2154

\bibitem[{Kastner {et~al.}(2002)Kastner, Huenemoerder, Schulz, Canizares, \&
  Weintraub}]{Kastner2002}
Kastner, J.~H., Huenemoerder, D.~P., Schulz, N.~S., Canizares, C.~R., \&
  Weintraub, D.~A. 2002, \apj, 567, 434

\bibitem[{Kirienko(1993)}]{Kirienko1993}
Kirienko, A.~B. 1993, \astl, 19, 11

\bibitem[{Koldoba {et~al.}(2008)Koldoba, Ustyugova, Romanova, \&
  Lovelace}]{Koldoba2008}
Koldoba, A.~V., Ustyugova, G.~V., Romanova, M.~M., \& Lovelace, R. V.~E. 2008,
  \mnras, 388, 357

\bibitem[{Lequeux(2005)}]{Lequeux2005}
Lequeux, J. 2005, {The Interstellar Medium}, Astronomy and Astrophysics Library
  (Berlin/Heidelberg: Springer-Verlag)

\bibitem[{Lesaffre(2002)}]{Lesaffre2002}
Lesaffre, P. 2002, PhD thesis, Universit{\'e} Paris VII

\bibitem[{Lesaffre {et~al.}(2004)Lesaffre, Chi{\`e}ze, Cabrit, \& Pineau~des
  For{\^e}ts}]{Lesaffre2004}
Lesaffre, P., Chi{\`e}ze, J.-P., Cabrit, S., \& Pineau~des For{\^e}ts, G. 2004,
  \aap, 427, 147

\bibitem[{Levermore(1996)}]{Levermore1996}
Levermore, C.~D. 1996, JSP, 83, 1021

\bibitem[{Lowrie {et~al.}(2001)Lowrie, Mihalas, \& Morel}]{Lowrie2001}
Lowrie, R.~B., Mihalas, D., \& Morel, J.~E. 2001, JQSRT, 69, 291

\bibitem[{Matsakos {et~al.}(2013)Matsakos, Chi{\`e}ze, Stehl{\'e},
  Gonz{\'a}lez, Ibgui, de~S{\'a}, Lanz, Orlando, Bonito, Argiroffi, Reale, \&
  Peres}]{Matsakos2013}
Matsakos, T., Chi{\`e}ze, J.-P., Stehl{\'e}, C., {et~al.} 2013, \aap, 557, A69

\bibitem[{Matsakos {et~al.}(2014)Matsakos, Chi{\`e}ze, Stehl{\'e},
  Gonz{\'a}lez, Ibgui, de~S{\'a}, Lanz, Orlando, Bonito, Argiroffi, Reale, \&
  Peres}]{Matsakos2014b}
Matsakos, T., Chi{\`e}ze, J.-P., Stehl{\'e}, C., {et~al.} 2014, Proceedings of
  the International Astronomical Union, 9, 66

\bibitem[{Mignone(2005)}]{Mignone2005}
Mignone, A. 2005, \apj, 626, 373

\bibitem[{Mihalas \& Mihalas(1984)}]{Mihalas1984}
Mihalas, D. \& Mihalas, B.~W. 1984, {Foundations of radiation hydrodynamics}
  (New York, Oxford University Press)

\bibitem[{Muzerolle {et~al.}(1998)Muzerolle, Calvet, \&
  Hartmann}]{Muzerolle1998}
Muzerolle, J., Calvet, N., \& Hartmann, L.~W. 1998, \apj, 492, 743

\bibitem[{{Opacity Project Team}(1995)}]{OPT1995}
{Opacity Project Team}. 1995, {The Opacity Project, Vol. 1 } (Institute of
  Physics Publications, Bristol, UK)

\bibitem[{Orlando {et~al.}(2013)Orlando, Bonito, Argiroffi, Reale, Peres,
  Miceli, Matsakos, Stehl{\'e}, Ibgui, de~S{\'a}, Chi{\`e}ze, \&
  Lanz}]{Orlando2013}
Orlando, S., Bonito, R., Argiroffi, C., {et~al.} 2013, \aap, 559, A127

\bibitem[{Orlando {et~al.}(2010)Orlando, Sacco, Argiroffi, Reale, Peres, \&
  Maggio}]{Orlando2010}
Orlando, S., Sacco, G.~G., Argiroffi, C., {et~al.} 2010, \aap, 510, A71

\bibitem[{Oxenius(1986)}]{Oxenius1988}
Oxenius, J. 1986, {Kinetic theory of particles and photons. Theoretical
  foundations of Non-LTE plasma spectroscopy} (Springer Series in
  Electrophysics, Berlin: Springer)

\bibitem[{Peres {et~al.}(1982)Peres, Rosner, Serio, \& Vaiana}]{Peres1982}
Peres, G., Rosner, R., Serio, S., \& Vaiana, G.~S. 1982, \apj, 252, 791

\bibitem[{Press {et~al.}(1994)Press, Teukolsky, Vetterling, \&
  Flannery}]{NRF77}
Press, W.~H., Teukolsky, S.~A., Vetterling, W.~T., \& Flannery, B.~P. 1994, in
  Fortran Numerical Recipes (Cambridge: University Press)

\bibitem[{Rammacher \& Ulmschneider(1992)}]{Rammacher1992}
Rammacher, W. \& Ulmschneider, P. 1992, \aap, 253, 586

\bibitem[{Robrade \& Schmitt(2007)}]{Robrade2007}
Robrade, J. \& Schmitt, J. H. M.~M. 2007, \aap, 473, 229

\bibitem[{Rodriguez {et~al.}(2018)Rodriguez, Espinosa, \&
  Miguel~Gil}]{Rodriguez2018}
Rodriguez, R., Espinosa, G., \& Miguel~Gil, J. 2018, Physical Review E, 98,
  033213

\bibitem[{Sacco {et~al.}(2008)Sacco, Argiroffi, Orlando, Maggio, Peres, \&
  Reale}]{Sacco2008}
Sacco, G.~G., Argiroffi, C., Orlando, S., {et~al.} 2008, \apj, 491, L17

\bibitem[{Sacco {et~al.}(2010)Sacco, Orlando, Argiroffi, Maggio, Peres, Reale,
  \& Curran}]{Sacco2010}
Sacco, G.~G., Orlando, S., Argiroffi, C., {et~al.} 2010, \aap, 522, A55

\bibitem[{S{\k{a}}dowski {et~al.}(2014)S{\k{a}}dowski, Narayan, McKinney, \&
  Tchekhovskoy}]{Sadowski2014}
S{\k{a}}dowski, A., Narayan, R., McKinney, J.~C., \& Tchekhovskoy, A. 2014,
  \mnras, 439, 503

\bibitem[{Schmitt {et~al.}(2005)Schmitt, Robrade, Ness, Favata, \&
  Stelzer}]{Schmitt2005}
Schmitt, J. H. M.~M., Robrade, J., Ness, J.~U., Favata, F., \& Stelzer, B.
  2005, \aap, 432, L35

\bibitem[{Schwarzschild(1948)}]{Schwarzschild1948}
Schwarzschild, M. 1948, \apj, 107, 1

\bibitem[{Siwak {et~al.}(2018)Siwak, Ogloza, Moffat, Matthews, Rucinski,
  Kallinger, Kuschnig, Cameron, Weiss, Rowe, Guenther, \& Sasselov}]{Siwak2018}
Siwak, M., Ogloza, W., Moffat, A. F.~J., {et~al.} 2018, \mnras, 478, 758

\bibitem[{Sobotka {et~al.}(2016)Sobotka, Heinzel, {\v S}vanda, Jur{\v c}{\'a}k,
  del Moro, \& Berrilli}]{Sobotka2016}
Sobotka, M., Heinzel, P., {\v S}vanda, M., {et~al.} 2016, \apj, 826, 49

\bibitem[{Spitzer(1998)}]{Spitzer1998}
Spitzer, L. 1998, {Physical Processes in the Interstellar Medium} (Wiley-VCH)

\bibitem[{Spitzer \& H{\"a}rm(1953)}]{Spitzer1953}
Spitzer, L. \& H{\"a}rm, R. 1953, PhRv, 89, 977

\bibitem[{Stehl{\'e} \& Chi{\`e}ze(2002)}]{Stehle2002}
Stehl{\'e}, C. \& Chi{\`e}ze, J.-P. 2002, in SF2A-2002: Semaine de
  l'Astrophysique Francaise, ed. F.~{Combes} \& D.~{Barret}, 493

\bibitem[{Stelzer \& Schmitt(2004)}]{Stelzer2004}
Stelzer, B. \& Schmitt, J. H. M.~M. 2004, \aap, 418, 687

\bibitem[{Ulmschneider {et~al.}(2005)Ulmschneider, Rammacher, Musielak, \&
  Kalkofen}]{Ulmschneider2005}
Ulmschneider, P., Rammacher, W., Musielak, Z.~E., \& Kalkofen, W. 2005, \apj,
  631, L155

\bibitem[{van Leer(1973)}]{vanLeer1973}
van Leer, B. 1973, in Proceedings of the Third International Conference on
  Numerical Methods in Fluid Mechanics, Vol. 1 (Springer, New York), 163--168

\bibitem[{Vernazza {et~al.}(1973)Vernazza, Avrett, \& Loeser}]{Vernazza1973}
Vernazza, J.~E., Avrett, E.~H., \& Loeser, R. 1973, \apj, 184, 605

\bibitem[{Verner \& Ferland(1996)}]{Verner1996a}
Verner, D.~A. \& Ferland, G.~J. 1996, \apjs, 103, 467

\bibitem[{Vidal {et~al.}(1995)Vidal, Matte, Casanova, \& Larroche}]{Vidal1995}
Vidal, F., Matte, J.~P., Casanova, M., \& Larroche, O. 1995, Phys. Plasmas, 2,
  1412

\bibitem[{von Neumann \& Richtmyer(1950)}]{vonNeumann1950}
von Neumann, J. \& Richtmyer, R.~D. 1950, J. Appl. Phys., 21, 232

\bibitem[{Voronov(1997)}]{Voronov1997}
Voronov, G.~S. 1997, Atomic Data and Nuclear Data Tables, 65, 1

\bibitem[{Walder \& Folini(1996)}]{Walder1996}
Walder, R. \& Folini, D. 1996, \aap, 315, 265

\bibitem[{Yan {et~al.}(1998)Yan, Sadeghpour, \& Dalgarno}]{Yan1998}
Yan, M., Sadeghpour, H.~R., \& Dalgarno, A. 1998, \apj, 496, 1044

\end{thebibliography}

	\begin{appendix}
	\section{Opacity tables\label{DESA_ref_app_opacities}}

		The specificity of the accretion shocks study led us to work on dedicated opacity tables. We expose in the appendix the reasons behind this choice and the creation process. The resulting opacity table is accessible upon request.

		\subsection{Motivation}

			\begin{figure}[!htb]
				\centering
				\includegraphics[scale=.8]{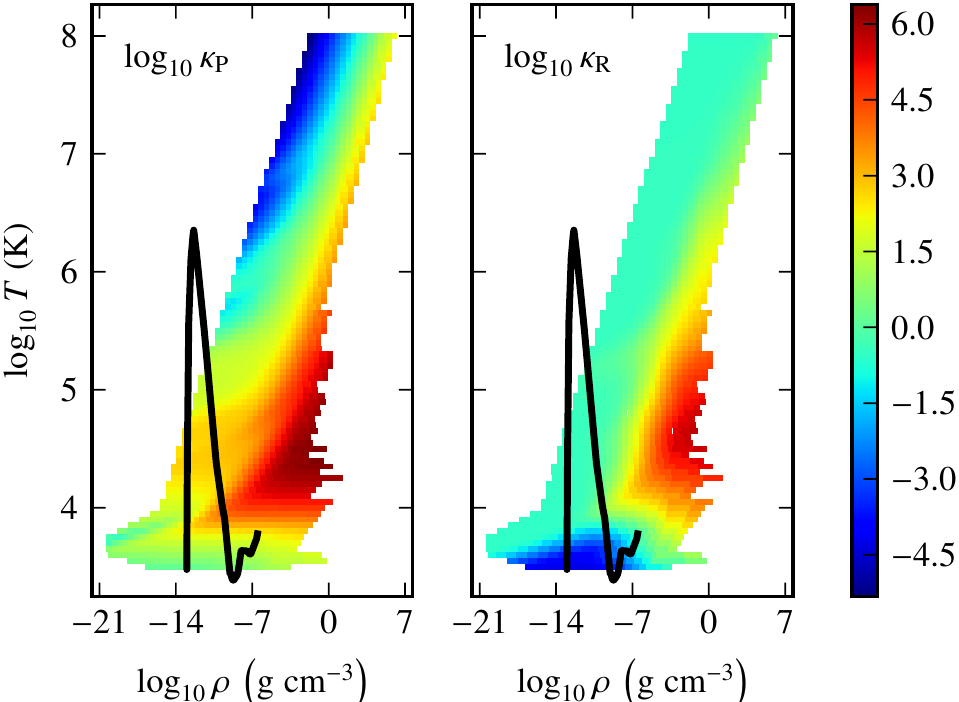}
				\caption{Planck (\texttt{left}) and Rosseland (\texttt{right}) opacities (in \si{\square\cm\per\gram}) with respect to gas density and temperature -- any in log scale, as provided by the Opacity Project \citep{OPT1995}. The \texttt{black curve} is a typical characteristic of an accretion column.\label{DESA_ref_OP}}
			\end{figure}

			\noindent Most available opacity tables are defined on a \emph{slanted} $(\rho,T)$ or $(n_e,T)$ domain (see e.g. Figure \ref{DESA_ref_OP}). However, one peculiarity of accretion shock structures is the presence of a low density hot post-shock plasma (black curve vertex in Figure \ref{DESA_ref_OP}) that explores a domain uncovered by publicly available tables. More complete tables are thus mandatory for the present study.

		\subsection{Choice of primary tables}

			\begin{figure}[!htb]
				\centering
				\includegraphics[scale=.8]{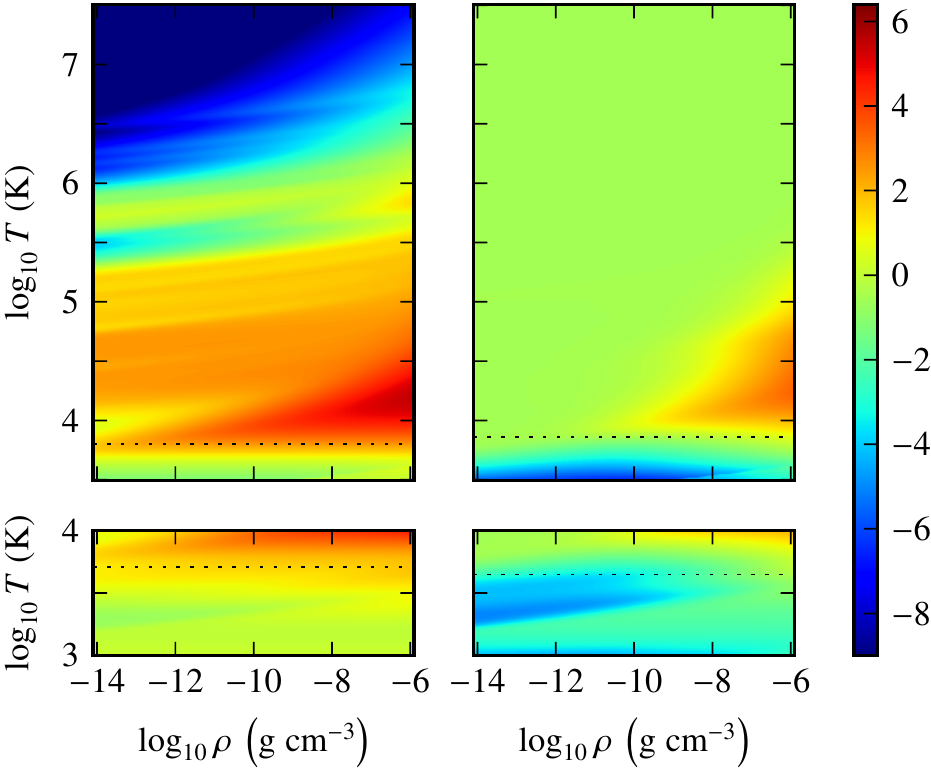}
				\caption{SYNSPEC (\texttt{top}) and Ferguson (\texttt{bottom}) Planck (\texttt{left}) and Rosseland (\texttt{right}) opacities (in \si{\square\cm\per\gram}) with respect to gas density and temperature -- any in log scale. The \texttt{dotted lines} show transition temperatures chosen for each table (see Annexe \ref{DESA_ref_app_Method}).\label{DESA_ref_CS+JF}}
			\end{figure}

			To cover the density and temperature range corresponding to our conditions, we implement in the code SYNSPEC (see Section \ref{DESA_ref_SYNSPEC}), initially dedicated to stellar atmospheres, modules allowing to generate LTE monochormatic opacities at a given density and temperature. These monochromatic opacities were then averaged with the proper weighting functions to generate the adequate Rosseland and Planck mean opacities tables (hereafter called "SYNSPEC tables", see Figure \ref{DESA_ref_CS+JF}, top panels). These opacities are consistent with Opacity Project \citep[see e.g.][]{OPT1995} data, that we use as reference, for $T$ between \SI{e3.5}{\K} and \SI{e7.5}{\K}. The advantage of SYNSPEC comes from the high number of atomic species considered, since a very detailed chemical composition is necessary to model the radiation properties of a plasma at high temperatures.\\

			However, below \SI{e3.5}{\K}, the molecular chemistry cannot be neglected, but is not included in this work on SYNSPEC. We completed thus the SYNSPEC tables with low temperature molecular opacities provided by \citet{Ferguson2005} between \SI{e3}{\K} and \SI{e4}{\K} ("Ferguson tables", see Figure \ref{DESA_ref_CS+JF}, bottom panels), that show excellent agreement with Opacity Project at upper temperatures. To facilitate the merging process, we obtained from the authors tables with compatible density and temperature grid (Ferguson, priv. comm.): mesh points from Ferguson and SYNSPEC tables are identical in the common domain (\SIrange{e3.5}{e4}{\K} and \SIrange{e-14}{e-6}{\gram\per\cubic\cm}).

		\subsection{Preliminary study}

			\subsubsection{Analysis of primary tables}

				Considering opacity variations as well as temperature and density ranges, we decided to work with the logarithm of all these quantities. As first derivatives, we use then:
				\begin{equation}\left\{\begin{array}{l@{\,}c@{\,}c@{\,}c@{\,}c@{}c}
					\partial_{lT}l\kappa &=& \dfrac{\partial\log_{10}\kappa}{\partial\log_{10}T} &=& \dfrac{T}{\kappa}&\dfrac{\partial\kappa}{\partial T}\\[13pt]
					\partial_{l\rho}l\kappa &=& \dfrac{\partial\log_{10}\kappa}{\partial\log_{10}\rho} &=& \dfrac{\rho}{\kappa}&\dfrac{\partial\kappa}{\partial\rho}
					\end{array}\right.
				\end{equation}
				where $\kappa$ stands for $\kappa_\mathrm{P}$ or $\kappa_\mathrm{R}$.

				\begin{figure}[!htb]
					\begin{minipage}[c]{.59\linewidth}
						\includegraphics[scale=.8]{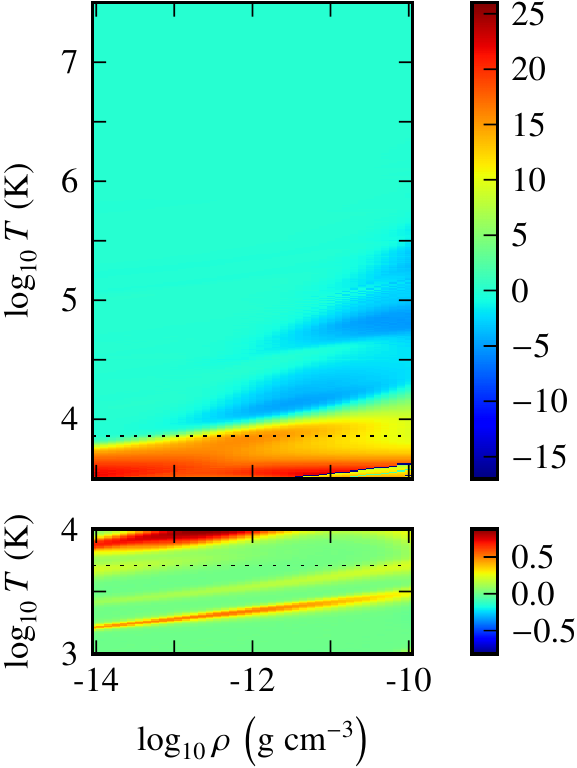}
					\end{minipage}\hfill \begin{minipage}[c]{.36\linewidth}
						\caption{$\partial_{lT}l\kappa_\mathrm{R}$ from SYNSPEC table (\texttt{top}) and $\partial_{l\rho}l\kappa_\mathrm{P}$ from Ferguson table (\texttt{bottom}) with respect to the logarithm of gas density and temperature. The \texttt{dotted lines} show transition temperatures chosen for each table. Some anomalies are revealed, especially around \SI{e3.7}{\K} and high densities: this zone is cutted during the merging process.\label{DESA_ref_dCSJF}}\vspace{.6cm\null}
					\end{minipage}
				\end{figure}

				Preliminary analysis of SYNSPEC and Ferguson tables revealed local aberrations, especially looking at the temperature or density derivatives (see Figure \ref{DESA_ref_dCSJF}). We may use the merging process to smooth most aberrations.

			\subsubsection{Physical and numerical constraints\label{DESA_ref_app_constraints}}

				In order to get a satisfying merging, several numerical and physical constraints must be respected:
				\begin{itemize}[nosep]
					\item as far as possible, opacities must be of class $C^1$ (values and first derivatives must be continuous);
					\item the transition region should be as narrow as possible;
					\item the transition region must encompass anomalies encountered in both primary tables.
				\end{itemize}

				Such a table is composed of a limited number of discrete points: the first constraint can be reported to the interpolation method as far as opacity values in the transition present smooth variations.\\
				To ensure a smooth transition between the molecular and the atomic (primary) tables, the transition must not take into consideration the values within the transition. The transition values loose then any physical meaning, and must be as few as possible\footnote{We note that the Ferguson tables showed opacity discontinuities in their hottest and densest part, as SYNSPEC tables in their coolest and densest part$^{(*)}$ (see Figure \ref{DESA_ref_dCSJF}). Since the values within the transition region are ignored, we use it to artificially remove anomalies: as far as possible, the transition region must be chosen so that it covers most of them.\\
				$^{(*)}$ Few anomalies remains in regions that are not explored in our simulations (see Figure \ref{DESA_ref_dFull}); this problem is postponed for now.}.

		\subsection{Merging process}

			\subsubsection{Method\label{DESA_ref_app_Method}}

				In this Section, the index "A" refers to values taken at the lower transition temperature, as the index "B" for the upper ones. The transition temperatures chosen to merge SYNSPEC and Ferguson tables are:
				\begin{itemize}[nosep]
					\item $T_\mathrm{\!A} = \SI{e3.71}{\K}$ and $T_\mathrm{\!B} = \SI{e3.80}{\K}$ for $\kappa_\mathrm{P}$ ($\sim \SI{1200}{\K}$ wide);
					\item $T_\mathrm{\!A} = \SI{e3.65}{\K}$ and $T_\mathrm{\!B} = \SI{e3.86}{\K}$ for $\kappa_\mathrm{R}$ ($\sim \SI{2800}{\K}$ wide).
				\end{itemize}

				The problem is decoupled in temperature and in density. First, we consider the merging at each mesh density as an isolated problem, and apply a correction -- if needed -- to improve smoothness along the density.

			\subsubsection{Merging along temperature}

				To satisfy the class $C^1$ constraint, we combined (see for instance \citealp{Auer2003} and \citealp{Ibgui2013}~\footnote{\citet{Fritsch1984} derivatives are used in these papers; they generalise van Leer slopes to non-regular grids.}):
				\begin{itemize}[nosep]
					\item piecewise cubic Hermite polynomials, which ensure continuity of values ($\kappa_\mathrm{A},\kappa_\mathrm{B}$) and derivatives ($\partial_{lT}l\kappa_\mathrm{A},\partial_{lT}l\kappa_\mathrm{B}$) at each transition limit;
					\item \citet{vanLeer1973} slopes to compute $\partial_{lT}l\kappa_\mathrm{A}$ and $\partial_{lT}l\kappa_\mathrm{B}$, so as to prevent the apparition of spurious extrema in forcing their location to the estimated closest mesh point.
				\end{itemize}

				For each grid density $\rho_{\!j}$, opacity at temperature $T_{\!i}\in[T_\mathrm{\!A},T_\mathrm{\!B}]$ is estimated using the formula:
				\begin{equation}
					\begin{array}[b]{@{}r@{}l@{}}
						\log_{10}\kappa(T_{\!i};\rho_{\!j}) = u_i^2\,(3-2u_i)\,\log_{10}\kappa_\mathrm{B} &+ (u_i-1)\hphantom{^2}u_i^2\,h\;\partial_{lT}l\kappa_\mathrm{B}\\[4pt]
							+ (u_i-1)^2\,(2u_i+1)\,\log_{10}\kappa_\mathrm{A} &+(u_i-1)^2\,u_i\,h\;\partial_{lT}l\kappa_\mathrm{A}
						\end{array}
				\end{equation}
				with $h = \log_{10}T_\mathrm{\!B} - \log_{10}T_\mathrm{\!A}$ and $u_i = (\log_{10}T_\mathrm{\!i}-\log_{10}T_\mathrm{\!A})/h$. This expression can be rewritten as a 3$^\text{rd}$ degree polynomial in $u_i$.

			\subsubsection{Density correction}

				At this stage, we reached class $C^1$ along temperature, but there is no guarantee of continuity along density. However, in practice, it was $C^1$, except for few mesh temperatures ${T_{\!i}}^*$.\\

				Since the dependency in density is held by the 3$^\text{rd}$ degree polynomial coefficients, we look at the behaviour of each of them with respect to density. Every coefficient showed spurious variations nowhere but at densities ${\rho_{\!j}}^*$. We apply then piecewise cubic Hermite polynomials along with \citeauthor{vanLeer1973} slopes (density derivatives) to estimate these coefficients for each ${\rho_{\!j}}^*$. These new coefficients are then used to reestimate opacity values along the temperature for the ${\rho_{\!j}}^*$.

			\subsubsection{Final tables -- interpolation process}

				\begin{figure}[!htb]
					\centering
					\includegraphics[scale=.8]{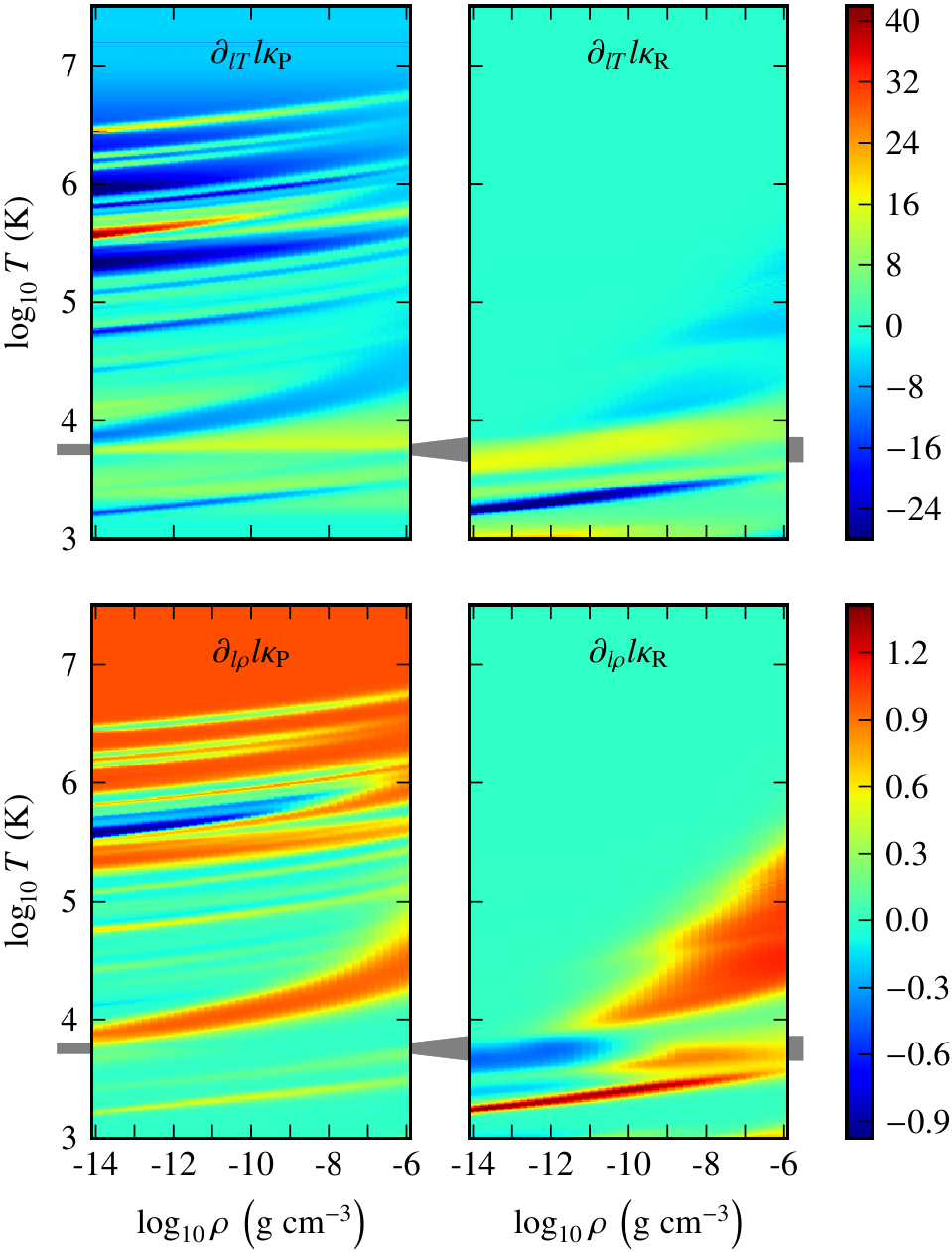}
					\caption{Merged table Planck (\texttt{left}) and Rosseland (\texttt{right}) opacity temperature (\texttt{top}) and density (\texttt{bottom}) first derivatives in the ($\rho$,$T$) plane, any in log scale; the \texttt{grey shape} represents the transition region.\label{DESA_ref_dFull}}
				\end{figure}

				We checked smoothness of the result by looking at the first derivatives. Figure \ref{DESA_ref_dFull} shows no anomaly within the transition temperature range $[T_\mathrm{\!A},T_\mathrm{\!B}]$ (grey shape). The remaining anomalies are not reached in our simulations.\\

				The interpolation process is copied from the merging method, i.e. piecewise cubic Hermite polynomials along with van Leer slopes, since it satisfies criteria described in Section \ref{DESA_ref_app_constraints}. Interpolation is first performed along temperature at the $2\times2$ grid densities framing the requested density, so as to calculate van Leer slopes at the requested temperature and interpolate along density.\\
				Interpolating along temperature and then density showed to be slightly more accurate than interpolation along density first. This is arguably due to stronger variations of opacities (especially Planck opacity) with respect to temperature.

	\section{Chromospheric model\label{DESA_ref_app_chromo}}

		One of our objectives is to describe the dynamics of the column and its impact on the chromosphere, as well as the feedback of the chromosphere on the column. This requires then to include an adequate description of the physical mechanism leading to the chromospheric heating. This appendix presents the simple but self-consistent model of a chromosphere used in this work.

		\subsection{Motivations and limits\label{DESA_ref_app_Motiv}}

			The study of the solar chromosphere is a tough problem in itself. Its modelling is of interest for us since the base of the accretion column lies in the stellar chromosphere: the dynamics and observability of the column base may then depend on its structure and dynamics. Moreover, the chromosphere may be heated locally by the accretion process. The inner heating mechanism in the chromosphere is still subject of debates: it is mainly thought to originate either from acoustic waves dissipation (\citealp{Biermann1946,Schwarzschild1948}; or more recently \citealp{Sobotka2016}) or from MHD waves dissipation \citep{Alfven1947,Jess2015}.

			Most accretion simulations model the stellar atmosphere -- when it is modelled -- as a hydrostatic plasma layer "tuned up" with ad-hoc sources to recover both temperature and pressure profiles \citep[see e.g. the heating function empirically introduced by][]{Peres1982}. Although this must work for a static structure, it is delicate to predict the dynamic behaviour of such a structure facing the continuous perturbation from an infalling plasma flow: such solution is not adapted to studies involving (in a self-consistent way) the dynamics of a perturbed atmosphere, like in the context of accretion.

			We do not pretend to develop a "state of the art" model in this paper: we only aim at using a reasonable model that is both dynamic and self-consistent with our radiation hydrodynamics model. In our 1D model, we do not consider any magnetic effect but a very effective confinement of the accretion flow along the field lines. To allow fast qualitative comparison between our model and theoretical models \& observations (see Figure \ref{DESA_ref_app_chromo-snap}), we only used solar parameters (i.e. abundances, luminosity, mass and radius).

		\subsection{Acoustic waves and shocks}

			Acoustic waves are generated by photospheric granulation \citep[see e.g.][]{Judge2006}. These waves propagate upwards up to the height where their velocity overcome the local sound speed, and degenerate then into shocks. The nature of this mechanism is random: two different locations at the stellar surface will be crossed over by acoustic shocks that ought to be out of phase one with each other.

\begin{figure}[htb]
				\centering
				\includegraphics[scale=.8]{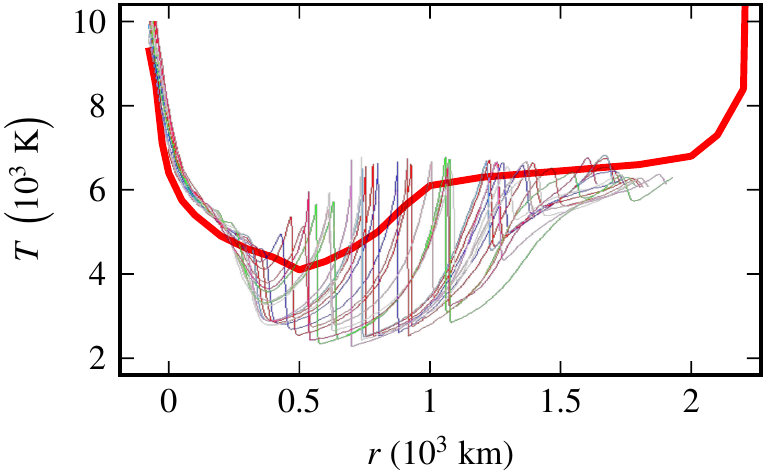}
				\caption{Successive snapshots of acoustic waves propagation (\texttt{thin lines}) and mean chromospheric temperature \citep[\texttt{thick red line},][]{Vernazza1973}. The simulation setup is described in Section \ref{DESA_ref_CL-Setup}; $r=\SI{0}{\km}$ locates the solar photosphere. Adapted from \citet{Chieze2012}.\label{DESA_ref_app_chromo-snap}}
			\end{figure}

			In our simulations, acoustic energy is supplied in the form of a monochromatic sinusoidal motion of the first Lagrangian interface \citep[$T = \SI{60}{\s}$ and $f_\mathrm{acc} = \SI{e8}{\erg\per\square\cm\per\s}$, see e.g.][]{Rammacher1992,Ulmschneider2005,Kalkofen2007}. Resulting acoustic waves propagate and degenerate into shocks. Figure \ref{DESA_ref_app_chromo-snap} shows several temperature snapshots of such simulation along with the chromospheric model from \citet{Vernazza1973}. Below \SI{300}{\km}, acoustic waves are damped and hardly appear on snapshots. Above \SI{500}{\km}, waves are fully degenerated into shocks: their strength is then governed by the balance between steepening in the pressure gradient and dissipation. Since the corona and the upper chromosphere (above \SI{e3}{\km}) are readily crushed by the accretion flow, the heating of these areas is not considered in our model.

		\subsection{From solar to stellar chromosphere\label{DESA_ref_app_chromoTTS}}

			Observations of the solar chromosphere provide time and space averages of thermodynamics quantities ($\rho$, $T$, $p$, $\ldots$). Detailed observation of CTTS chromospheres would demand higher space and time resolution than the ones permitted by current observational technologies. Most works on this field rely then on scaling laws \citep[see e.g.][]{Ayres1979,Calvet1983} or ad hoc fittings to recover specific observational features \citep[see e.g.][]{Dumont1973,Cram1979,Calvet1984,Batalha1993}.

	\section{Radiation source terms in the Hybrid model \label{DESA_ref_app_radsourceshybrid}}

		This work encompasses several radiation regimes, from optically thick LTE radiation transfer (Section \ref{DESA_ref_LTEtransfer}) to optically thin coronal NLTE regime (Section \ref{DESA_ref_Lambda}). The momenta equations (Section \ref{DESA_ref_RT}) can handle all of them, assuming the proper radiation source terms are provided.

		In the LTE case, both radiation energy and momentum source terms are well defined (Eq. \eqref{DESA_ref_sourcesLTE}). In coronal regime, this is not the case. Gas and radiation are decoupled in such a regime. Radiation only acts then as a gas energy sink: the radiation energy source term (the gas sink) boils down to a cooling function \citep[see e.g.]{Kirienko1993}. Computing the radiation flux is irrelevant in such regime and then \emph{no radiation momentum} source term is provided. That is why we set $\vec{\mathfrak{s}}^\dagger_{M_\mathrm{r}}$ to $\vec{0}$.

		In the "Hybrid" setup, we aim at modelling radiative conditions that are neither LTE nor coronal regimes but something in between. To determine if the situation is closer to one or the other, and how close, we choose to look at the probability for a photon to escape the accretion column (see Eq. \eqref{DESA_ref_eqzeta}). We use it as a weighting factor to average the source terms, as shown in Section \ref{DESA_ref_Zeta}.\\
		The process is straightforward for the radiation energy source term, but not for the radiation momentum source term since $\vec{\mathfrak{s}}^\dagger_{M_\mathrm{r}}$ remains unknown. We assume then that the coronal Rosseland mean opacity may not significantly differ from its LTE value. This intuition is reinforced by preliminary calculations concerning NLTE radiative collisional opacities (Pérez, priv. com.).

	\end{appendix}

\end{document}